\documentclass[twocolumn]{aastex63}%
\usepackage{lineno}
\usepackage{amsmath}
\usepackage{mathrsfs}
\usepackage{verbatim}

\submitjournal{ApJL}
\shorttitle{Interacting Kilonovae in AGN Disks}
\shortauthors{Ren et al.}
\begin{document}
\title{Interacting Kilonovae: Long-lasting Electromagnetic Counterparts to Binary Mergers in the Accretion Disks of Active Galactic Nuclei}

\correspondingauthor{Zi-Gao Dai}
\email{daizg@ustc.edu.cn}

\author[0000-0002-9037-8642]{Jia Ren}
\affiliation{School of Astronomy and Space Science, Nanjing University, Nanjing 210023, China}
\affiliation{Key Laboratory of Modern Astronomy and Astrophysics (Nanjing University), Ministry of Education, Nanjing 210023, China}
\author[0000-0001-8955-0452]{Ken Chen}
\affiliation{School of Astronomy and Space Science, Nanjing University, Nanjing 210023, China}
\affiliation{Key Laboratory of Modern Astronomy and Astrophysics (Nanjing University), Ministry of Education, Nanjing 210023, China}
\author[0000-0002-8385-7848]{Yun Wang}
\affiliation{Key Laboratory of Dark Matter and Space Astronomy, Purple Mountain Observatory, Chinese Academy of Sciences, Nanjing 210034, China}
\affiliation{Department of Astronomy, School of Physical Sciences, University of Science and Technology of China, Hefei 230026, China}
\author[0000-0002-7835-8585]{Zi-Gao Dai}
\affiliation{Department of Astronomy, School of Physical Sciences, University of Science and Technology of China, Hefei 230026, China}
\affiliation{School of Astronomy and Space Science, Nanjing University, Nanjing 210023, China}

\begin{abstract}
We investigate the dynamics and electromagnetic (EM) signatures of
neutron star-neutron star (NS-NS) or neutron star-black hole (NS-BH) merger ejecta
that occurs in the accretion disk of an active galactic nucleus (AGN).
We find that the interaction between ejecta and disk gas leads to important effects
on the dynamics and radiation. We show five stages of  the ejecta dynamics:
gravitational slowing down, coasting, Sedov-Taylor deceleration in the disk,
re-acceleration after the breakout from the disk surface,
and momentum-conserved snowplow phase.
Meanwhile, the radiation from the ejecta is so bright that its typical peak luminosity reaches
a few times $10^{43}-10^{44}~\rm erg~s^{-1}$.
Since most of the radiation energy has converted from the kinetic energy of merger ejecta,
we call such an explosive phenomenon an interacting kilonova (IKN).
It should be emphasized that IKNe are very promising,
bright EM counterparts to NS-NS/BH-NS merger events in AGN disks.
The bright peak luminosity and long rising time
(i.e., ten to twenty days in UV bands, thirty to fifty days in optical bands,
and one hundred days to hundreds of days in IR bands)
allow most survey telescopes to have ample time to detect an IKN.
However, the peak brightness, peak time,
and evolution pattern of the light curve of an IKN are similar to
a superluminous supernova in a galactic nucleus and a tidal disruption event
making it difficult to distinguish between them.
But it also suggests that IKNe might have been present in recorded AGN transients.
\end{abstract}

\section{Introduction}
Binary compact star mergers, especially black hole-black hole (BH-BH),
black hole-neutron star (BH-NS), and NS-NS mergers,
are meaningful gravitational wave (GW) radiation sources for the GW detectors
\citep{Abbott_2016_Abbott_prl_v116.p61102..61102,
Abbott_2017_Abbott_prl_v119.p30..33,
Abbott_2019_Abbott_PhysicalReviewX_v9.p31040..31040,
Abbott_2021_Abbott_PhysicalReviewX_v11.p21053..21053}.
Although the formation channels for such binaries are still under debate,
two main channels are widely discussed, i.e.,
isolated binary evolution (e.g.,
\citealp{Belczynski_2010_Bulik_apj_v714.p1217..1226,
deMink_2016_Mandel_mnras_v460.p3545..3553,
Santoliquido_2020_Mapelli_apj_v898.p152..152})
and dense environments causing dynamical interactions
(e.g., globular clusters,
\citealp{Sigurdsson_1993_Hernquist_Nature_v364.p423..425,
Rodriguez_2015_Morscher_prl_v115.p51101..51101};
galactic nuclei, \citealp{Antonini_2016_Rasio_apj_v831.p187..187,
Fragione_2019_Leigh_mnras_v488.p2825..2835};
and active galactic nucleus (AGN) disks,
\citealp{McKernan_2012_Ford_mnras_v425.p460..469,
McKernan_2014_Ford_mnras_v441.p900..909,
Bartos_2017_Kocsis_apj_v835.p165..165,
Stone_2017_Metzger_mnras_v464.p946..954}).
It was suggested that the GW kicked remnant BH after a BH-BH merger in  the AGN disk
can cause an ultraviolet (UV)-optical flare
\citep{McKernan_2019_Ford_apj_v884.p50..50L}.
It was reported that a plausible optical EM counterpart ZTF19abanrhr
\citep{Graham_2020_Ford_prl_v124.p251102..251102}
for the high-mass BH-BH merger event GW190521
\citep{Abbott_2020_Abbott_prl_v125.p101102..101102}
can be explained in this model (but also see
\citealp{Ashton_2020_Ackley_ClassicalandQuantumGravity_v38.p235004..235004},
\citealp{Nitz_2021_Capano_apj_v907.p9..9L}, and
\citealp{Pan_2021_Yang_apj_v923.p173..173}).

In the same manner as BH-BH binaries, BH-NS/NS-NS binaries can also coalesce in the disks of AGNs.
AGN disks may contain a variety of compact stars and their binaries,
including compact stars and their binaries that are captured by the disk
(e.g., \citealp{Stone_2017_Metzger_mnras_v464.p946..954,
Fabj_2020_Nasim_mnras_v499.p2608..2616});
directly formed massive stars that subsequently collapse in the disk
(e.g., \citealp{Goodman_2003__mnras_v339.p937..948});
and efficient migration of compact stars
(e.g., \citealp{Bellovary_2016_Low_apj_v819.p17..17L}).
The formation and merger of binary systems could be accelerated
by dense environments and a large number of compact stars
(e.g., \citealp{Tagawa_2020_Haiman_apj_v898.p25..25,
Samsing_2022_Bartos_nat_v603.p237..240}).
For example,
\cite{McKernan_2020_Ford_mnras_v498.p4088..4094}
showed that
if a fraction $f_{\rm AGN}\sim 0.1$ of BH-BH mergers with the observed rate
${R_{\rm BH-BH} \sim 100~\rm{Gpc}^{-3}~\rm{yr}^{-1}}$
comes from AGNs
\citep{McKernan_2018_Ford_apj_v866.p66..66},
the observed rate of BH-NS mergers from the AGN channel is
${R_{\rm BH-NS} \sim f_{\rm AGN}\left[10,300\right] ~\rm{Gpc}^{-3}~\rm{yr}^{-1}}$,
and the NS-NS rate is
${R_{\rm NS-NS} \sim f_{\rm AGN}\left[0.2,400\right] ~\rm{Gpc}^{-3}~\rm{yr}^{-1}}$,
respectively\footnote{
\cite{Perna_2021_Tagawa_apj_v915.p10..10}
also showed a different NS-NS merger rate:
${R_{\rm NS-NS}\sim \left[0.1,5\right] ~\rm{Gpc}^{-3}~\rm{yr}^{-1}}$.
The findings of \cite{Tagawa_2021_Kocsis_apj_v908.p194..194}
concerning much lower merger rates of binary compact objects
consisting of at least one NS were presented in Table~3 in their paper.}.
\cite{McKernan_2018_Ford_apj_v866.p66..66}
was also shown that the fraction of BH-NS mergers with the BH-NS mass ratio
$q\equiv M_{\rm BH}/M_{\rm NS}\lesssim 5$ is approximately $\sim 0.11$.
In this case, parts of the NS will be disintegrated as ejecta and thrown from the binary system
\citep[e.g.,][]{Shibata_2009_Kyutoku_prd_v79.p44030..44030,
Bartos_2013_Brady_ClassicalandQuantumGravity_v30.p123001..123001}.
Thus, the detected rate of BH-NS/NS-NS mergers with remnant ejecta is
${0.1R_{\rm BH-NS}+R_{\rm NS-NS} \sim f_{\rm AGN}\left[1.2,430\right] ~\rm{Gpc}^{-3}~\rm{yr}^{-1}}$.

For the BH-NS/NS-NS mergers that have remnant ejecta,
$r$-process elements could power the so-called ``kilonova'' transients
\citep{Li_1998_Paczynski_apj_v507.p59..62L,
Metzger_2010_MartinezPi_mnras_v406.p2650..2662}.
Previous research has focused on kilonovae that
are associated with typical environments of short gamma-ray bursts,
i.e., low circumburst medium density and located most likely in a galactic halo
(e.g., \citealp{Berger_2013_Fong_apj_v774.p2011..2014,
Tanvir_2013_Levan_Nature_v500.p547..549,
Jin_2016_Hotokezaka_NatureCommunications_v7.p..,
Jin_2020_Covino_NatureAstronomy_v4.p77..82}).
A typical kilonova, e.g., AT~2017gfo, with peak luminosity
$\sim 10^{41}~\rm{erg}~\rm{s}^{-1}$
and duration time $\sim 10$ days,
is easy to observe in such environments
(e.g., \citealp{Coulter_2017_Foley_Science_v358.p1556..1558,
Cowperthwaite_2017_Berger_apj_v848.p17..17L,
Kasen_2017_Metzger_Nature_v551.p80..84}).
But, as another important place for merger generation,
kilonovae that occur in AGNs have not been well investigated.
Dense, dusty, chaotic environments in the AGN disk
provide an excellent place to study the properties of kilonovae
or other transients in such complex environments.

The AGN disk environment has recently been the subject of several studies
investigating stellar transients and their electromagnetic (EM) emission.
The first case is the gamma-ray burst (GRB).
The emission of jets has been studied under the precondition of
whether the progenitor has opened a cavity or not.
For the non-cavity scenario,
\cite{Perna_2021_Lazzati_apjl_v906.p7..7L} looked at the dynamics and radiation of the relativistic jet
in the disk and studied the typical radii and observability of both the prompt and afterglow radiation;
\cite{Wang_2022_Lazzati_mnras_v.p..} simulated the diffused GRB afterglows from the disk of an AGN.
Upon reprocessing by disk, the GRB afterglows became dimmer but lasted longer.
Additionally, the detection of such transients in the optical bands is significantly more promising;
\cite{Zhu_2021_Zhang_apj_v906.p11..11L} studied the breakout of a cocoon
produced by a choked jet within the AGN disk
and found that the breakout emission can be detected at $z\lesssim 0.5$ using the
Einstein Probe on the X-ray band.
In their analysis, they argued that typical kilonovae in the AGN environment are dim and difficult to detect,
but they ignored the energy injected from the interaction between ejecta and disk gas.
\cite{Yuan_2022_Murase_apj_v932.p80..80}
has studied the radiation of jets if a cavity has formed before the jet,
finding that up-scattered photons from the disk cavity could be detected in the GeV to the sub-TeV range.
The second case is the stellar explosion/outflow in the disk.
\cite{Zhu_2021_Yang_apjl_v914.p19..19L}
studied thermonuclear explosions and accretion-induced collapses of white dwarfs in the AGN disk.
Moreover, they noted that the interaction between disk gas and explosion ejecta
has a significant impact on radiation.
\cite{Grishin_2021_Bobrick_mnras_v507.p156..174}
analyzed and simulated supernovae explosion within or offset from AGN disks.
The explosions triggered by supernovae are capable of producing transient events
of high luminosity up to $10^{44}-10^{45}~\rm erg~s^{-1}$ in such environments.

As mentioned above, we note that the interaction between ejecta and disk gas
can play an important role in the dynamics and radiation of merger ejecta.
We investigate the dynamics and radiation properties
of an embedded merger remnant in the AGN disk in this paper.
Our results indicate that an interaction between merger ejecta
and gas in the AGN disk may lead to an interesting type of optical transient in the galactic nuclei.
This paper is organized as follows.
In Section~{\ref{model}} we present our model.
In Section~{\ref{results}} we describe numerical results in detail.
We discuss some additional considerations in Section~{\ref{discuss}}.
We summarize the significance of our results in Section~{\ref{Summary}}.

\section{The Model}\label{model}
Numerical simulations predicted that compact binary mergers could occur
in migration traps of the AGN disks.
\cite{Bellovary_2016_Low_apj_v819.p17..17L}
obtained the typical radius of mergers as $\sim 20-300r_{\rm s}$
to a central supermassive black hole (SMBH),
where $r_{\rm s}=2GM_{\rm SMBH}/c^2$ is the Schwarzschild radius of the SMBH
with $G$ being the gravitational constant and $c$ being the speed of light.
\cite{Tagawa_2020_Haiman_apj_v898.p25..25}
obtained a more distant location at $\sim 10^3-10^4r_{\rm s}$.
In this work, we are mainly concerned with the mergers that occur  at $\sim 10^2-10^4r_{\rm s}$.

\subsection{The AGN disk}
Near the migration traps, the density of the disk midplane is $\geqslant 10^{-10}~\rm g\; cm^{-3}$ \citep{Kato_1998_Fukue__v.p..,Wang_2021_Liu_apjl_v911.p14..14L},
which is enough to influence the dynamics and radiation behavior of the merger ejecta.
Therefore, it is necessary to calculate the disk properties with careful consideration.
We use the SG disk model which assumes that
the outer part disk is heated sufficiently to maintain marginal gravitational stability,
presumably by massive stars formed within the disk
\citep{Sirko_2003_Goodman_mnras_v341.p501..508,Yang_2019_Bartos_apj_v876.p122..122}.
We use this model and assume the vertically isothermal approximation of the accretion disk \citep{Kato_1998_Fukue__v.p..}
which gives a gaussian distribution of vertical profile to obtain the required disk parameters at each radius, e.g.,
the gas density $\rho$, pressure $P$, half-height of the disk $H$, and gas opacity $\kappa$.

\subsection{Spatial Distribution of Ejecta}\label{Sec:Ejecta Distribution}
A series of numerical simulations revealed the complex spatial
and temporal distribution of ejecta from merged NS-NS and NS-BH binaries
(e.g., \citealp{Dessart_2009_Ott_apj_v690.p1681..1705,
Perego_2014_Rosswog_mnras_v443.p3134..3156,
Radice_2016_Galeazzi_mnras_v460.p3255..3271,
Bovard_2017_Martin_prd_v96.p124005..124005,
Fujibayashi_2018_Kiuchi_apj_v860.p64..64,
Fernandez_2019_Tchekhovsk_mnras_v482.p3373..3393,
Christie_2019_Lalakos_mnras_v490.p4811..4825}).
We ignore the possible long-lasting activity of the central remnant.
Since the timescale of ejecta propagation in the AGN disk is much longer than
the timescale of an outflow (e.g.,
\citealp{Fujibayashi_2018_Kiuchi_apj_v860.p64..64,
Fernandez_2019_Tchekhovsk_mnras_v482.p3373..3393}),
one just needs to consider the spatial distribution of ejecta.
The ejecta has three components:
dynamic ejecta, wind, and secular ejecta,
whose geometry and composition have distributions depending on the polar angle and velocity
(e.g., \citealp{Metzger_2017__LivingReviewsinRelativity_v20.p1..59,
Rosswog_2022_Korobkin_arXiveprints_v.p2208..14026arXiv}).
As \cite{Perego_2017_Radice_apj_v850.p37..37L} proposed,
the angular distribution of dynamic ejecta is well approximated by $F(\theta) = \sin^2\theta$,
the wind geometry favors a rather uniform distribution in polar angle $F(\theta) \approx \rm const$
for $\theta \lesssim \pi/3$ \citep{Martin_2015_Perego_apj_v813.p2..2}, and
the flow of secular ejecta is dominantly equatorial
$F(\theta) = \sin^2\theta$.
Since the mass fraction of the wind component to the total mass of ejecta is very low,
we assume an equatorial-dominated outflow of the ejecta
$M_{\rm ej}(\theta)\propto \sin^2\theta$.
We ignore the radial evolution of velocity in this study for simplicity
and assume that materials move in a specific direction as thin shells.
Using numerical simulation results,
we determine the Root-Mean-Square (RMS) radial velocities of ejecta with different polar angles.\footnote{
As shown in Section~\ref{dynamics}, the ejecta rapidly evolved into a Sedov-Taylor stage.
Even though the radial profile of the ejecta may appear complex at first,
it will become smeared after this stage.
Generally, the interior of a Sedov blast wave is composed of very hot, low-density material
\citep[e.g.,][Figure~9]{Reynolds_2017__HandbookofSupernovae_v.p1981..2004}.
A majority of the mass and kinetic energy is contained within the outermost 10\% of the blast radius,
and the pressure drops slightly behind the shock then be quasi-uniform.
Therefore, the thin-shell approximation and subsequent modeling approach we adopted are roughly accurate.}
We set
$v_{\rm ej}(\theta) = v_{\rm po}+(v_{\rm eq}-v_{\rm po})\sin^2\theta$,
where $v_{\rm po}$ ($v_{\rm eq}$) is the ejecta velocity at $\theta=0$~($\theta=\pi/2$).

\subsection{Ejecta and Disk Gas Interacting Model}\label{Sec:model}
During the NS-NS/NS-BH merger,
a compact object is left at the remnant center, meanwhile, the ejecta has expanded.
We use a thin shell model to describe the condition of ejecta after the initial expansion.
In our analysis, we assume central symmetry expansion for the ejecta,
with a symmetric axis perpendicular to the AGN disk.
An approximate three-dimensional model of ejecta expansion can be constructed
using the spherical coordinate frame~$(r,\theta,\varphi)$.
The origin of coordinates is set to be  the remnant center initially.
In this work, we ignore the proper motion
of the central compact remnant, the gravity of SMBH,
and the effect of the shear motion of the AGN disk.
In our calculations, the ejecta shell is divided into small patches,
and convection between neighboring patches has not taken into account.
The distance between a patch at the direction of $(\theta,\varphi)$
and the midplane of the AGN disk can be calculated by
$H_{\rm ej}(\theta,\varphi)=R_{\rm ej}(\theta,\varphi)\cos\theta$,
where $R_{\rm ej}(\theta,\varphi)$ is the distance to the remnant compact star.
The following equations hide ``$(\theta,\varphi)$'' from view,
but one should note that all calculations must be performed in an element
with solid angle $d\Omega= \sin \theta d\theta d\varphi$.

The density and pressure of the disk corresponding to the position of an element are
\citep{Kato_1998_Fukue__v.p..}
\begin{equation}
\rho_{\rm disk}=
\rho_0 \exp \left(-\frac{H_{\rm ej}^2}{2 H^2}\right),
\end{equation}
and
\begin{equation}
P_{\rm disk}=
P_0 \exp \left(-\frac{H_{\rm ej}^2}{2 H^2}\right),
\end{equation}
where $\rho_0$ and $P_0$ are the density and pressure in the midplane of AGN disk, respectively.
The density above the AGN disk drops rapidly
and eventually reaches the mean density of the nucleus sphere region
$\rho_{\rm nuc}\approx 10^{-17}~\rm g\; cm^{-3}$.
As the ejecta propagates, the disk parameters (e.g., $\rho_0$ and $P_0$) will change,
which is taken into consideration in our calculation.

First, the internal energy $U$ of the ejecta shell element located in the
$(\theta,\varphi)$-direction evolves as
\begin{equation}\label{dU}
\frac{d U}{d t}=
L_{\rm in}+L_{\rm sh}-L_{\rm e}+
\left(P_{\rm disk}-P_{\rm in}\right) \frac{dV}{dt},
\end{equation}
where $L_{\rm in}$ is the radioactive energy injection rate,
$L_{\rm sh}$ is the shock heating rate,
$L_{\rm e}$ is the radiation luminosity, and
$P_{\rm in}=(\hat{\gamma}-1)U/V$ is the pressure of ejecta
with $\hat{\gamma}$ represents the adiabatic index, respectively.
In this work we set $\hat{\gamma}=4/3$.
Note that $V=1/3 R_{\rm ej} S$ is the volume of an element and
$S=R_{\rm ej}^2d\Omega$.

The radioactive energy injection rate is expressed as
\begin{equation}
L_{\rm in}=M_{\rm ej,0}\dot{q}_r \eta_{\rm th},
\end{equation}
where $M_{\rm ej,0}$ is the mass of initial $r$-process ejecta,
the radioactive power per unit mass is
\citep{Korobkin_2012_Rosswog_mnras_v426.p1940..1949,
Barnes_2016_Kasen_apj_v829.p110..110,Metzger_2019__LRR_v23.p..}
\begin{equation}
\dot{q}_{r}
=4\times 10^{18}\left[
\frac{1}{2}-\frac{1}{\pi}\arctan
\left(\frac{t-t_{0}}{\sigma}\right)
\right]^{1.3} {\rm erg} \;{\rm s}^{-1} \; {\rm g}^{-1},
\end{equation}
where $t_0=1.3$~s, $\sigma=0.11$~s, and
the thermalization efficiency reads
\begin{equation}
\eta_{\rm th}=
0.36\left[
\exp \left(-0.56 t_{\rm day}\right)+
\frac{\ln \left(1+0.34 t_{\rm day}^{0.74}\right)}
{0.34 t_{\rm day}^{0.74}}
\right],
\end{equation}
with $t_{\rm day}=t/{\rm day}$.

The shock heating rate is (e.g., \citealp{Moriya_2013_Maeda_mnras_v435.p1520..1535})
\begin{equation}
L_{\rm sh}
=\frac{\epsilon}{2}\frac{dM}{dt}v_{\rm ej}^2
=\frac{\epsilon}{2} \rho_{\rm disk} v_{\rm ej}^3 S,
\end{equation}
where $\epsilon \in (0,1)$ is the efficiency of shock heating,
and $v_{\rm ej}$ is the velocity of an element.
We ignore the local kinematic velocity of the gas in the disk
in the above equation.
In this work, we set $\epsilon=0.9$,
since the swept mass is much higher than the initial mass of merger ejecta
\citep{Moriya_2013_Blinnikov_mnras_v428.p1020..1035}.
The radiation luminosity can be written as
\begin{equation}
L_{\rm e}=\frac{Uc}{R_{\rm ej}}
\frac{1-e^{-\tau}}{\tau}.
\end{equation}
Taking into account the impact of the disk environment,
one should consider the contribution to $\tau$ made by
the initial $r$-process ejecta $\tau_{\rm ej}$,
the swept disk gas $\tau_{\rm sw}$,
and the residual disk gas $\tau_{\rm disk}$, respectively.
Assuming that the radiation eventually exits in the direction perpendicular to the disk surface,
the total optical depth is
\begin{equation}
\begin{aligned}
\tau
& =
\tau_{\rm ej}+\tau_{\rm sw}+\tau_{\rm disk} \\
& \approx  \frac{\kappa_{\rm ej}M_{\rm ej,0}+\kappa_{\rm sw}M_{\rm sw}}{S}
+ \int_{H_{\rm ej}}^{\infty} \kappa_{\rm disk} \rho_0 \exp \left(-\frac{z^2}{2 H^2}\right) dz,
 \end{aligned}
\end{equation}
where $M_{\rm sw}$ is the mass of swept-up disk gas,
and $\kappa_{\rm ej}$, $\kappa_{\rm sw}$ and $\kappa_{\rm disk}$
are the opacities of $r$-process ejecta, shocked disk gas and disk gas, respectively.
We assume that $\kappa_{\rm sw}=0.34~\rm g^{-1}~cm^2$
and $\kappa_{\rm disk}$ can be calculated in equations of
\cite{Yang_2019_Bartos_apj_v876.p122..122}\footnote{
In order to obtain more realistic radiation properties, one needs to consider a more complex opacity table,
i.e., $\kappa(E_\gamma,T,Z,\rho)$, where $E_\gamma$ represents the photon energy
and $T,Z,\rho$ represents temperature, atomic number, and density of medium, respectively.
As a substitute, we use the average opacity which has been widely used in studies of
supernovae and accretion disk radiation processes.
In a larger scale, dust and gas from galactic nuclei may reprocess the emission,
but we have ignored this effect here.
The effect of this process can, however, be approximated by adding an extinction parameter, $A_\lambda$.}.
As a result of the extremely high density of disk gas,
diffused photons of ejecta in the disk are scattered in all directions,
which makes the outgoing photons almost isotropic \citep{Wang_2022_Lazzati_mnras_v.p..}.
We also assume that the radiation of the ejecta that breaks away from the disk is isotropic,
i.e., ignoring the effect of line of sight.
Next, we give the expression of the dynamics based on conservation of energy.
The total energy of an element in the ejecta-compact remnant system is
\begin{equation}\label{E}
 E=
 \frac{1}{2} M_{\rm ej}v_{\rm ej}^2
 + U
 - \frac{GM_{\bullet} M_{\rm ej}}{R_{\rm ej}},
\end{equation}
where $M_{\rm ej}=M_{\rm ej,0}+M_{\rm sw}$, and $M_{\bullet}$ is the mass of central compact remnant.
The variation of total energy is
\begin{equation}\label{dE}
dE=\left(L_{\rm in}-L_{\rm e}\right)dt-GM_{\bullet}\frac{dM_{\rm ej}}{R_{\rm ej}},
\end{equation}
The shock heating energy and the increase in gravitational potential energy  of the ejecta
are converted from its kinetic energy. Therefore, they have no impact on the system's total energy.
Substituting Equation~(\ref{dU}) and (\ref{E}) into Equation~(\ref{dE}),
one can get
\begin{equation}\label{v_ej>0}
\frac{dv_{\rm ej}}{dt} =
\frac{1}{M_{\rm ej}
v_{\rm ej}}\left[
\left(P_{\rm in}-P_{\rm disk}\right)\frac{dV}{dt}
-\frac{1+\epsilon}{2}v_{\rm ej}^2\frac{dM_{\rm ej}}{dt}\right]
-\frac{GM_{\bullet}}{R_{\rm ej}^2}.
\end{equation}
The first, second, and third terms on the right-hand side
are referred to as pressure, swept gas, and gravity, respectively.
With the above equations,
the dynamics and EM radiation can be solved simultaneously.
Once the equations of an element at $(\theta,\varphi)$-direction are solved,
the blackbody luminosity of an element at frequency $\nu$ reads
\begin{equation}
L_{\nu}=\frac{2 \pi h \nu^3}{c^2} \frac{1}{\exp \left(h \nu / k_{\rm B} T_{\rm eff}\right)-1}
R_{\rm ph}^2 d \Omega,
\end{equation}
where $R_{\rm ph}$ is the photosphere radius that can be written by
\begin{equation}
\int_{R_{\rm ph}}^{\infty} \kappa_{\rm disk} \rho_0 \exp \left(-\frac{z^2}{2 H^2}\right) dz =1,
\end{equation}
when $\tau_{\rm disk}>1$; if $\tau_{\rm disk}\leqslant1$, we set $R_{\rm ph}=R_{\rm ej}$ for simplicity
\citep{Yu_2018_Liu_apj_v861.p114..114,
Ren_2019_Lin_apj_v885.p60..60,
Qi_2022_Liu_apj_v925.p43..43}.
Then, the effective temperature $T_{\rm eff}$ can be calculated by
\begin{equation}
T_{\rm eff}=\left(\frac{L_{\rm e}}{\sigma_{\rm SB} R_{\rm ph}^2 d\Omega}\right)^{1/4}.
\end{equation}
The total EM luminosity of all elements can be obtained by
\begin{equation}
L_{\rm e,tot}=\int L_{\rm e}d\Omega.
\end{equation}
Similarly, the total EM luminosity at frequency $\nu$ is
\begin{equation}
L_{\nu, \rm tot}=\int L_{\nu}d\Omega.
\end{equation}

\section{Basic Numerical Results}\label{results}
Based on the assumption that the progenitor system does not open a cavity within the disk,
we present our numerical results of ejecta dynamics and radiation properties in this section.
Our calculations are based on the following parameters:
\begin{itemize}
\item
The parameters of the AGN disk are taken:
the viscosity parameter $\alpha=0.01$,
the radiation efficiency of disk $\eta=L_{\bullet,\rm Edd}/\dot{M}_{\bullet,\rm Edd}c^2=0.1$,
and the accretion rate $\dot{M}_{\bullet}=0.1\dot{M}_{\bullet,\rm Edd}$,
where $L_{\bullet,\rm Edd}$ and $\dot{M}_{\bullet,\rm Edd}$
are the Eddington luminosity and the corresponding accretion rate of the SMBH, respectively.
\item
For the ejecta dynamics described in Section~{\ref{dynamics}},
we set the central SMBH mass $M_{\bullet}=10^7M_{\odot}$,
and the merger formed at $10^3r_{\rm s}$ of the AGN disk.
To emphasize the main concern about dynamics,
we have ignored the change of disk parameters during ejecta propagation.
\item
For the radiation properties described in Section~{\ref{light_curves}},
the central SMBH masses are set as
$M_{\bullet}=10^6,~10^7$, and $10^8M_{\odot}$,
and the merger-forming radii are assumed to be
$10^2,~10^3$, and $10^4r_{\rm s}$ in the AGN disk for each SMBH mass, respectively.
In addition, the change in disk parameters during ejecta propagation is taken into consideration.
\item
The parameters of the merger remnant are taken:
the central remnant BH mass $M_{\bullet}=5M_{\odot}$,
the total initial ejecta mass $M_{\rm ej,0,tot}=6\times 10^{-2}M_{\odot}$,
and the initial radius of ejecta $R_{\rm ej,0}=10^9~\rm cm$.
For simplicity, the gray opacity of $r$-process ejecta is taken to be
 $\kappa_{\rm ej}=10 ~\rm cm^2~g^{-1}$.
We assume that the initial velocities of ejecta meet the form
$v_{\rm ej}(\theta) = v_{\rm po}+(v_{\rm eq}-v_{\rm po})\sin^2\theta$,
we set $v_{\rm po}=0.05c$, and $v_{\rm eq}=0.2c$, respectively
\citep{Zhu_2020_Yang_apj_v897.p20..20}.
The kinetic energy of the ejecta is $\sim 1.6\times 10^{51}$~erg,
similar with a typical SN.
\end{itemize}

\begin{figure*}
\centering
\includegraphics[width=0.32\hsize]{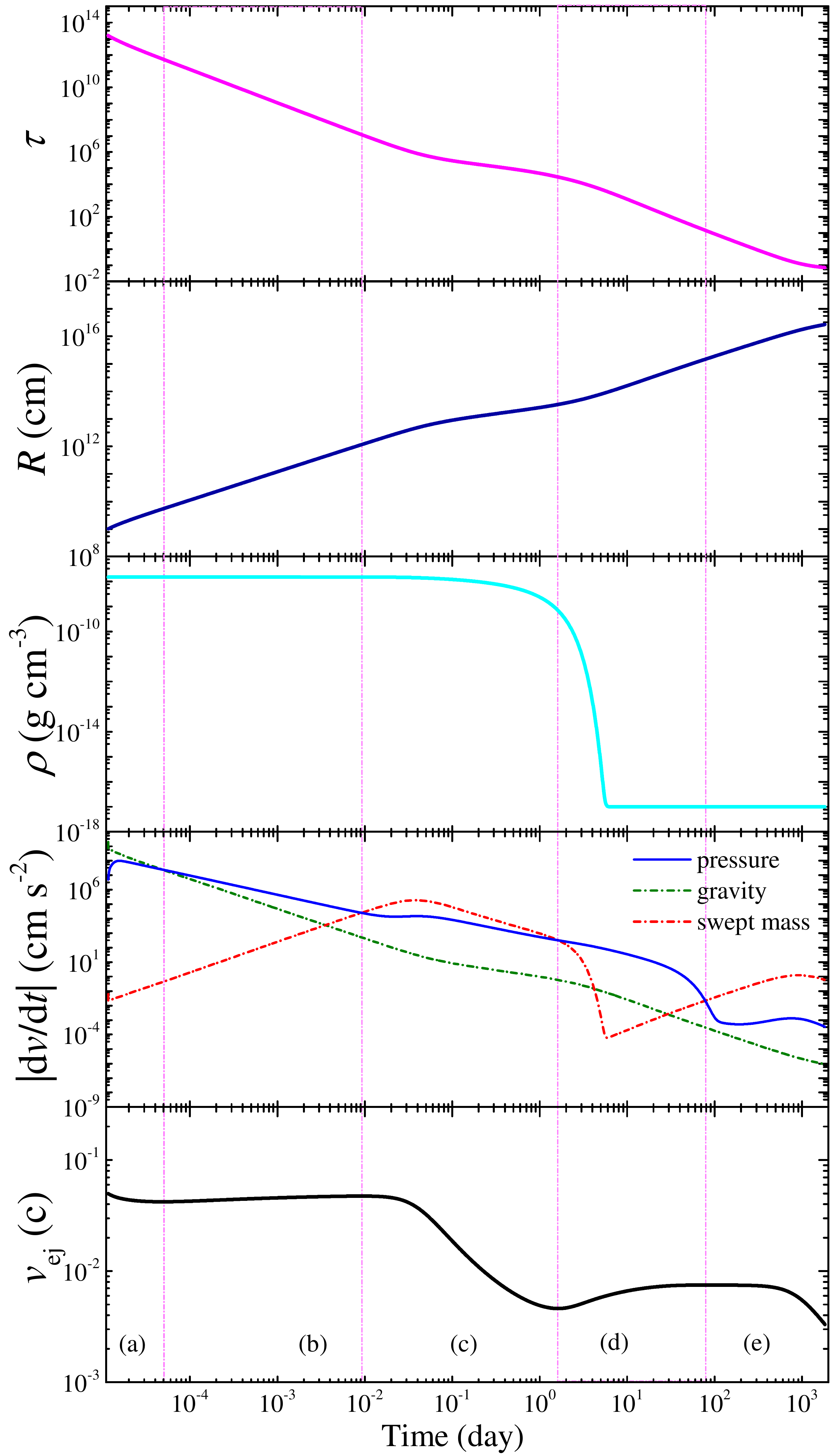}
\includegraphics[width=0.32\hsize]{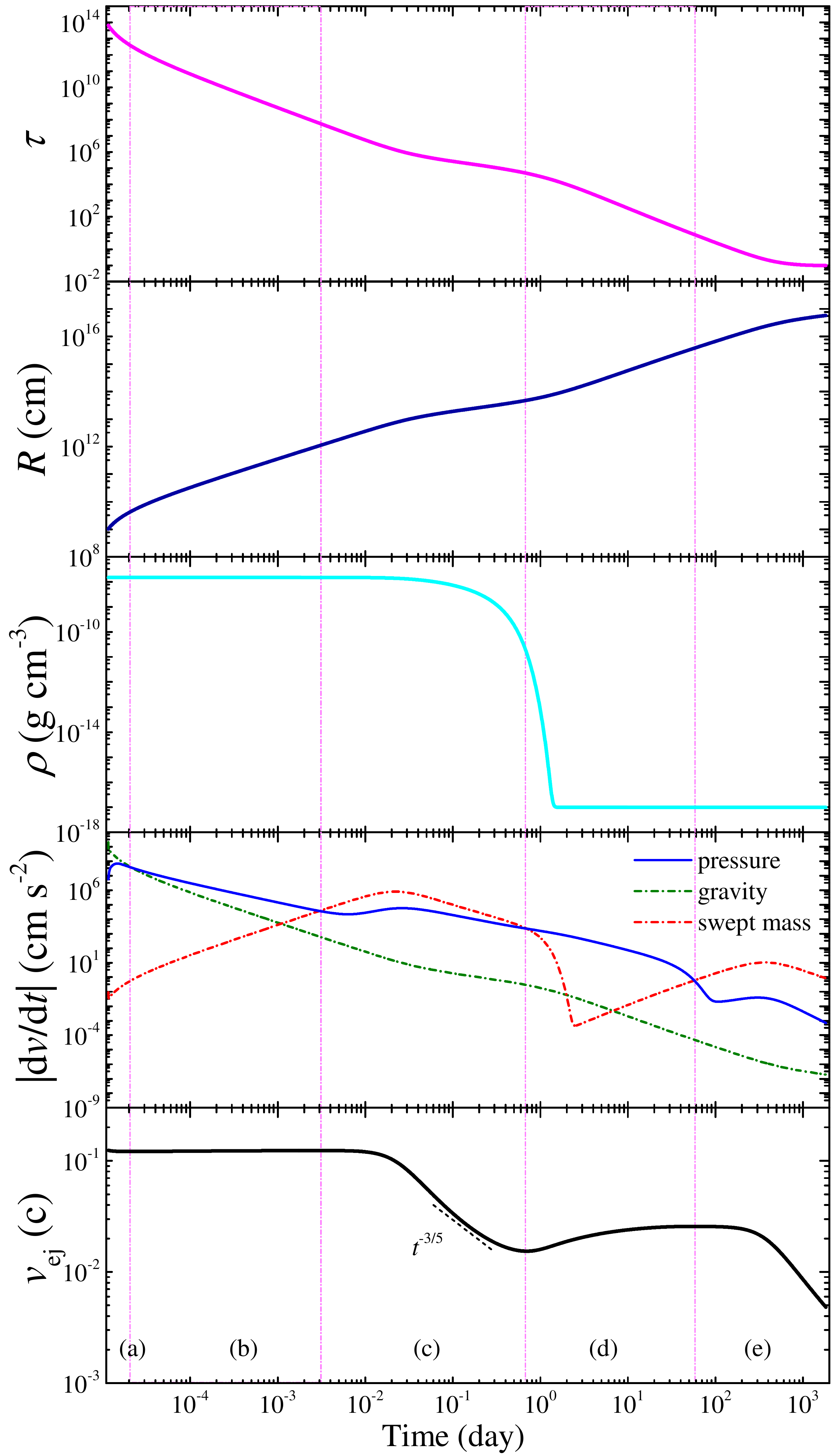}
\includegraphics[width=0.32\hsize]{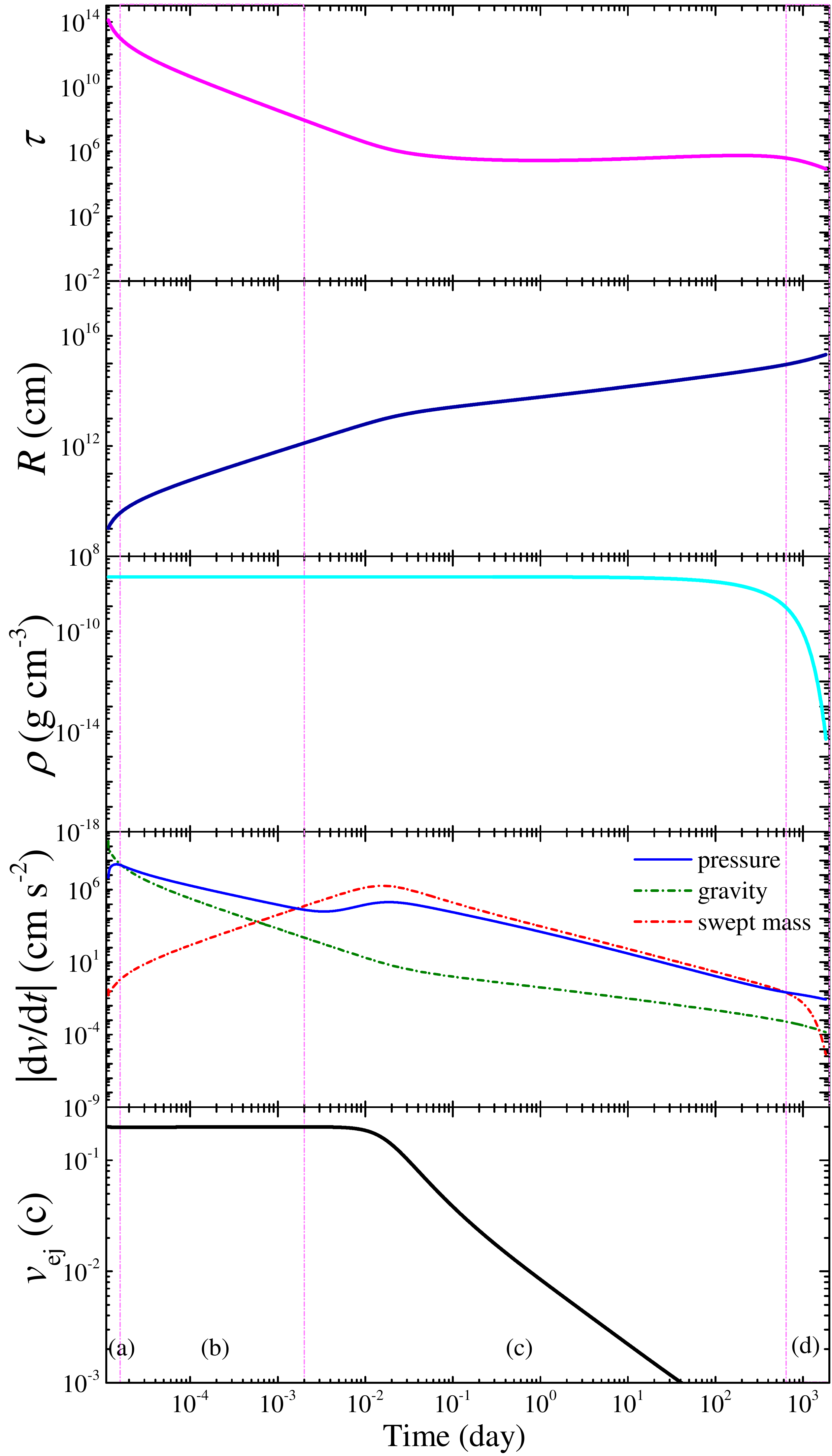}
\caption{
From top to bottom, the optical depth, the propagated distance of the ejecta,
the disk density in the position of the ejecta, the accelerated velocities,
and the dynamics of the ejecta evolve, respectively.
Three panels from left to right-hand sides show the
dynamics and other properties of ejecta at $0^{\circ}$, $45^{\circ}$, and $88^{\circ}$ directions, respectively.
The labels (a) - (e) mark five stages of the ejecta dynamics that are separated by vertical dash-doted lines.
The blue solid, red dash-doted, and green dash-doted lines are the accelerated velocities
from the pressure, the swept gas, and the gravity of the central compact remnant, respectively.
Note that the pressure accelerates the ejecta and the last two terms decelerate the ejecta.
The merger occurs in the disk radius as
$10^3~r_{\rm s}$ with the mass of SMBH being $10^7M_{\odot}$.
To emphasize the main concern about dynamics,
we have ignored the change of disk parameters,
e.g., $\rho_0$ and $P_0$, during ejecta propagation.
}
\label{MyFig1}
\end{figure*}
\subsection{Dynamics}\label{dynamics}
Figure~{\ref{MyFig1}} illustrates the dynamics of the shell of ejecta for
$\theta=0^{\circ}$, $45^{\circ}$, and $88^{\circ}$ as typical examples.
Interestingly, one can observe that the dynamics of ejecta propagating within the AGN disk are unusual.
It is easy to find that dynamic evolution can be divided into five distinct stages.
We label them in this figure as regions (a) - (e).
This means that different regions are defined
based on the transition between the dominant acceleration terms,
i.e., the three acceleration terms on the right-hand side of Equation~({\ref{v_ej>0}}).
We then discuss the physical reason for evolution of each phase.
\begin{itemize}
\item[(a)]  \textit{Gravitational slowing down:}
At the very beginning of the ejecta expansion in the deep AGN disk,
most of the ejecta energy is kinetic.
At this time, the ejecta is in a state of free expansion, slowed only by gravity.
\item[(b)] \textit{Free expansion:}
After stage (a), the deceleration of the gravitational term is negligible, and the ejecta is in a free expansion phase.
The so-called ``Sedov'' radius
$R_{\rm Sedov}=(3M_{\rm ej,0}/4\pi \rho_{\rm disk})^{1/3}\approx 7.8\times 10^{12}$~cm, for
$M_{\rm ej,0}=10^{-2}M_\odot$ and $\rho_{\rm disk}=10^{-8}~\rm g~cm^{-3}$,
which marks the end of the free expansion phase,
is determined by the condition that the swept-mass equals to the initial ejecta mass.
As a result of the heating effect of the interaction reverse shock,
the flat pressure structure is formed around the Sedov radius around time
$t_{\rm sw}=R_{\rm Sedov}/v_{\rm ej}=0.03~\rm d$, for $v_{\rm ej}=0.1c$.
Then, the ejecta reaches the self-similar evolution stage.
\item[(c)] \textit{Sedov-Taylor stage:}
Since the optical depth in the disk is very high,
as shown by the magenta line in the top panel of Figure~{\ref{MyFig1}},
the ejecta reaches the so-called ``Sedov-Taylor'' (ST) stage after the initial free expansion,
thus the velocity evolved as $t^{-3/5}$ (e.g.,
\citealp{Ostriker_1988_McKee_ReviewsofModernPhysics_v60.p1..68,
Haid_2016_Walch_mnras_v460.p2962..2978}).
We show the slope with a gray dashed line in Figure~{\ref{MyFig1}}.
By the end of this stage, due to propagation,
the ejecta is close to the vertical boundary of the disk.
Based on the rapid decrease in gas density, the deceleration tapers off and the ejecta is
re-accelerated by its inner pressure.
A clear correlation between the disk density and the ejecta velocity
can be observed in Figure~{\ref{MyFig1}}.
\item[(d)]  \textit{Re-accelerate after the breakout:}
In this stage, the ejecta is re-accelerated by the inner pressure
since the optical depth $\tau\sim\kappa M_{\rm ej}/R_{\rm ej}^2d\Omega$ is still much larger than unity.
We can obtain similar results with the re-acceleration of ejecta as in Figure~1 of
\citet{Grishin_2021_Bobrick_mnras_v507.p156..174}.
Note that $\tau\propto R_{\rm ej}^{-2}$,
because after the ejecta breaks out from the disk surface,
$M_{\rm ej}$ has a nearly constant value for some time thereafter.
Therefore, when the optical depth approaches unity,
the ejecta becomes optically thin and the radiative losses of energy increase,
leading to the termination of the adiabatic expansion of ejecta.
\item[(e)] \textit{Momentum conservation phase:}
During the last stage of the ejecta dynamics, the
thermal pressure in the ejecta decreases, and the expansion slows down.
It enters momentum conservation
between the ejecta and the environment gas at the galactic nucleus sphere region,
known as the ``snowplow phase''.
\end{itemize}

Different ejecta moving in different $\theta$ directions have different masses and velocities.
Consequently, there are different dynamics of ejecta evolving in different directions.
Based on the picture above, one can imagine that the ejecta in the vertical disk direction
has an extremely short ST phase and, due to its fast dropped optical depth,
enters the snowplow phase very quickly.
Meanwhile, the ejecta in the disk direction
has an extremely long ST phase and, due to its higher optical depth,
enters the snowplow phase at a very late time.
One can clearly observe these phenomena in Figure~{\ref{MyFig1}}.

\subsection{Luminosity Light Curves}\label{light_curves}
\begin{figure*}
\centering
\includegraphics[width=0.32\hsize]{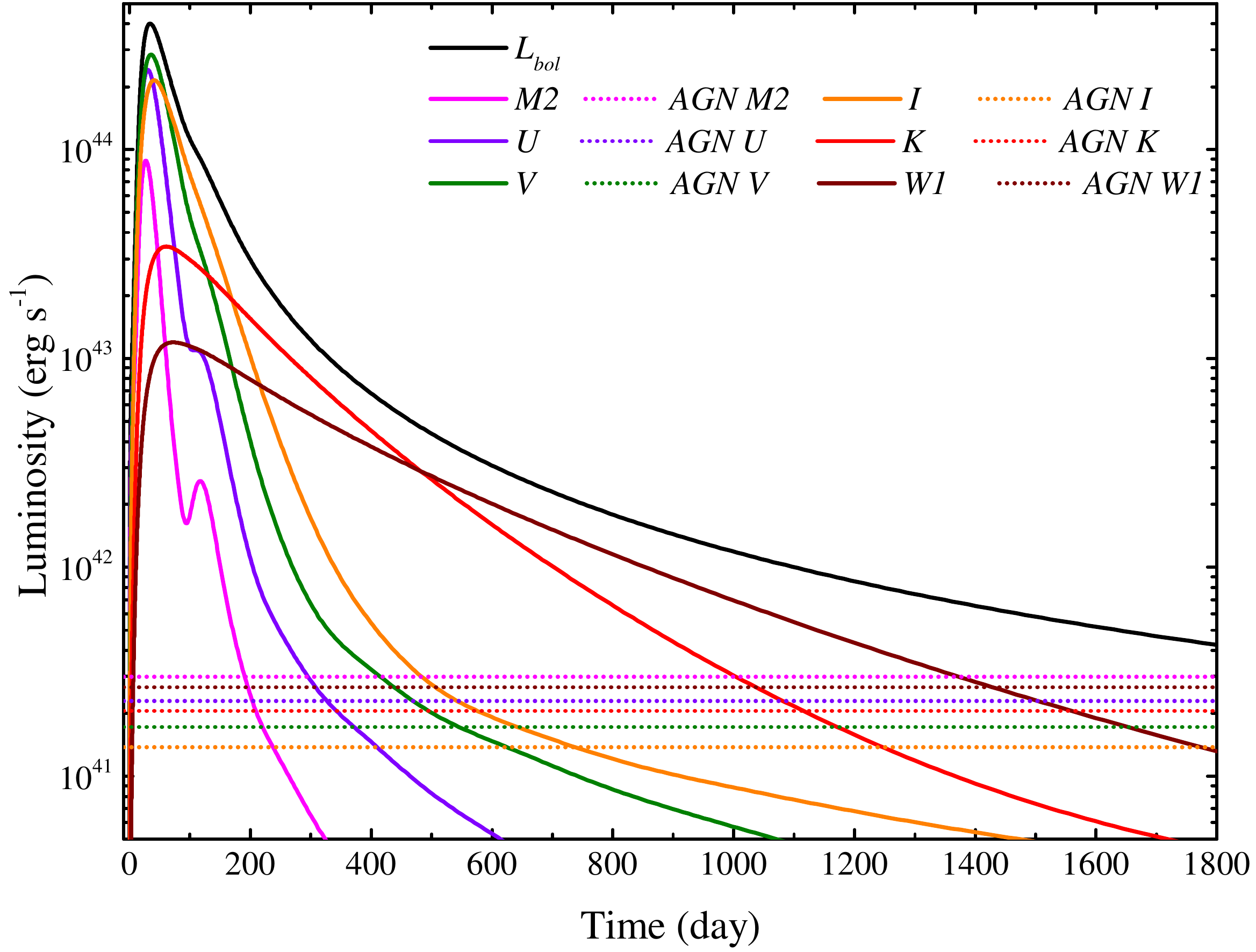}
\includegraphics[width=0.32\hsize]{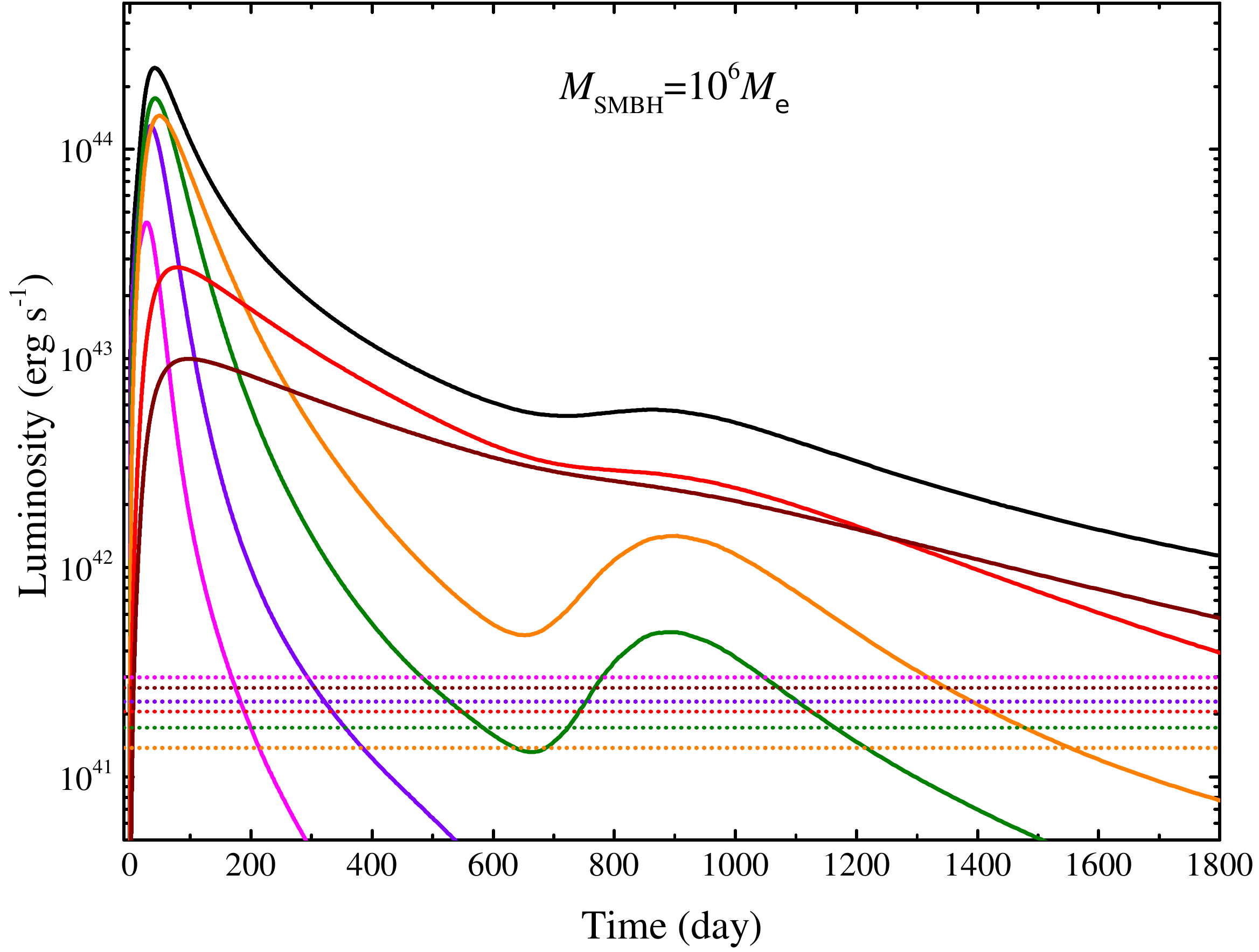}
\includegraphics[width=0.32\hsize]{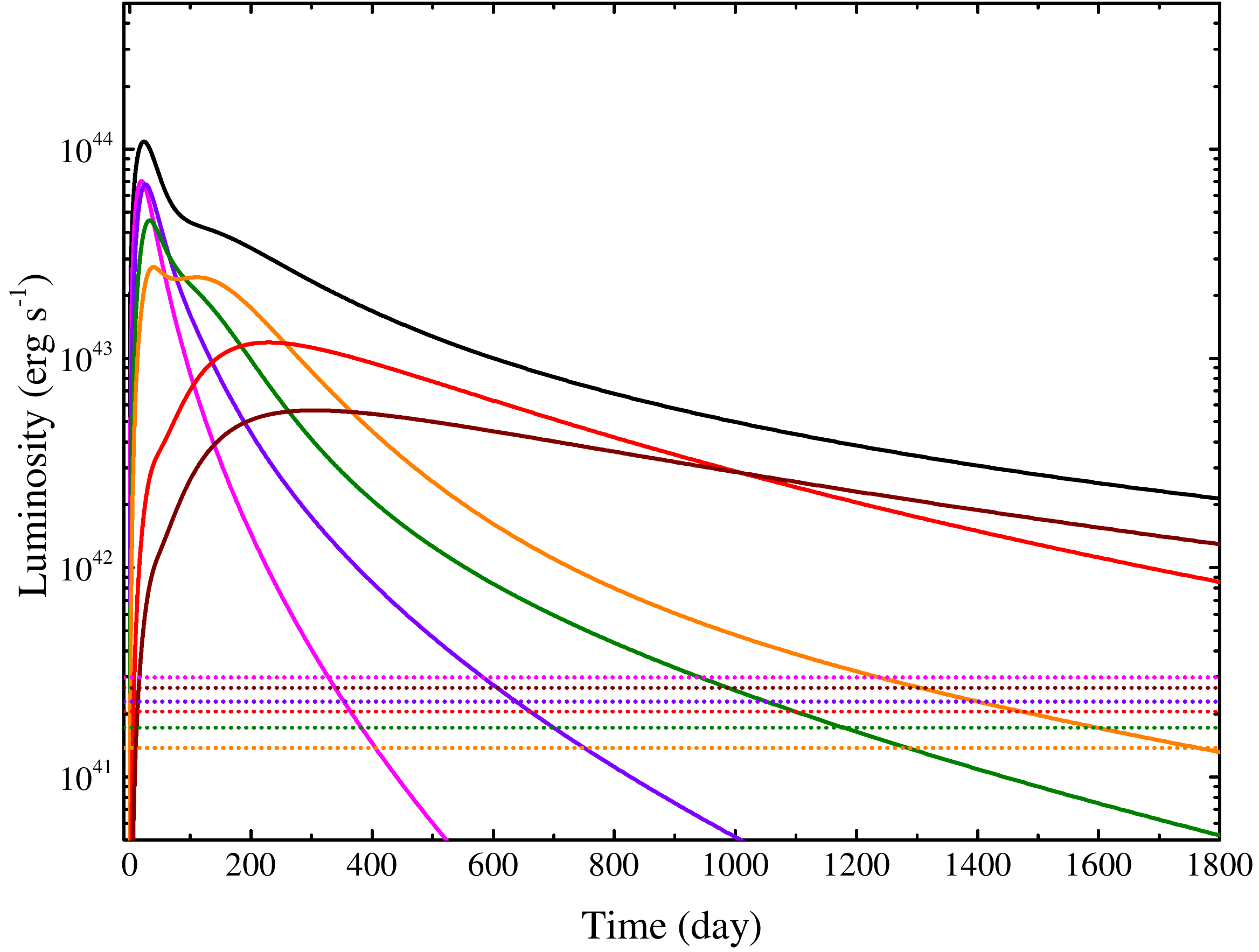} \\
\includegraphics[width=0.32\hsize]{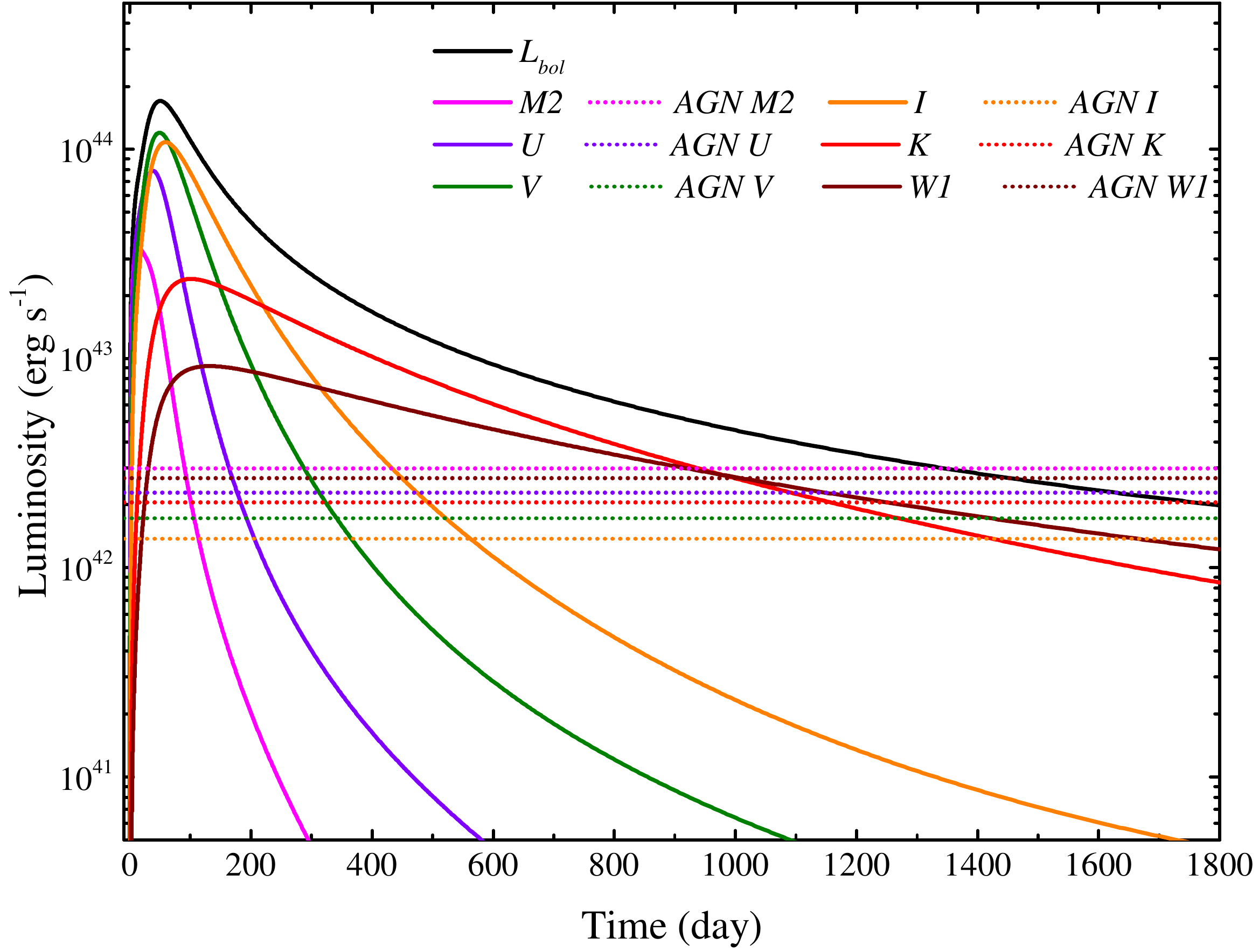}
\includegraphics[width=0.32\hsize]{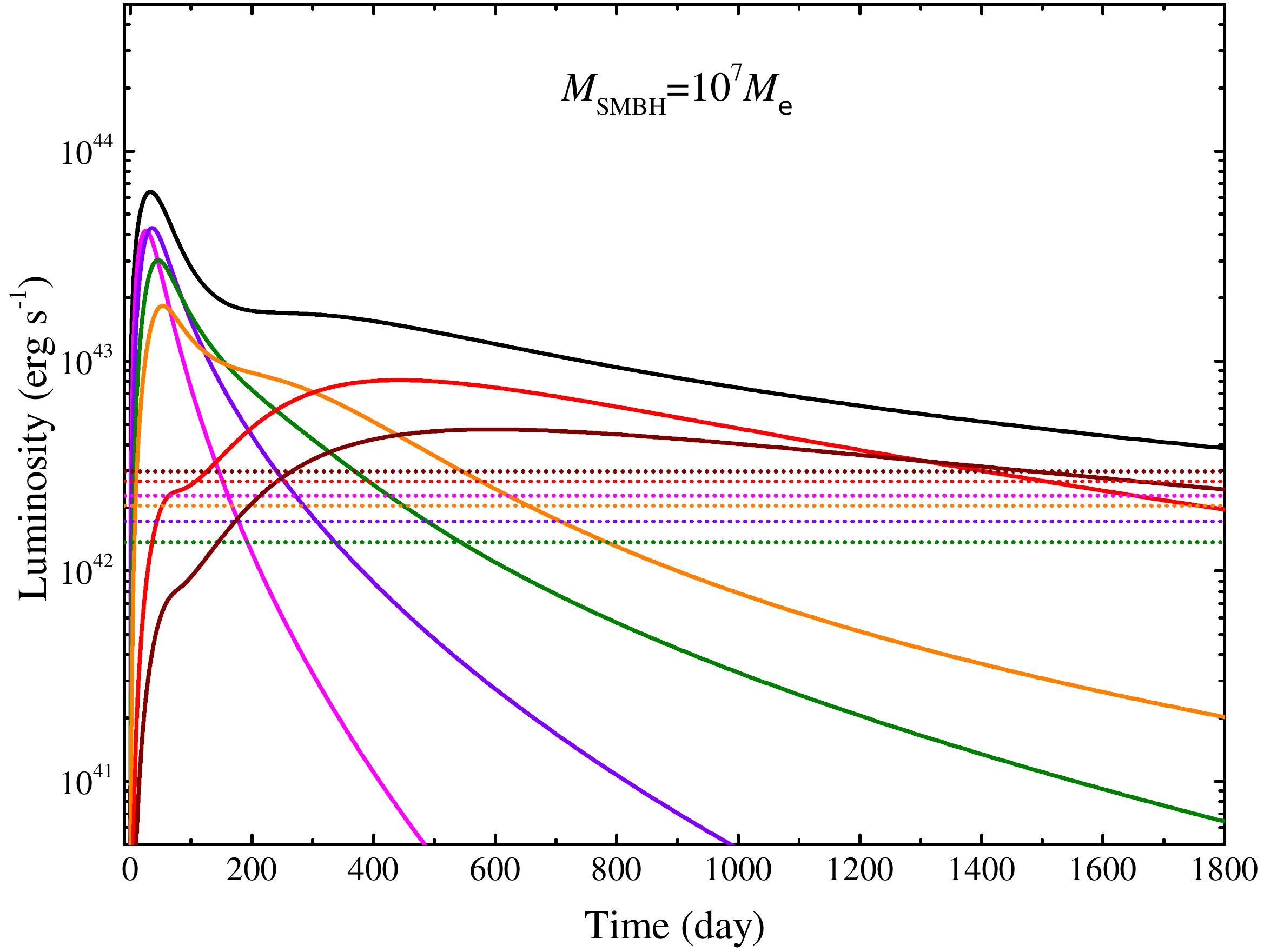}
\includegraphics[width=0.32\hsize]{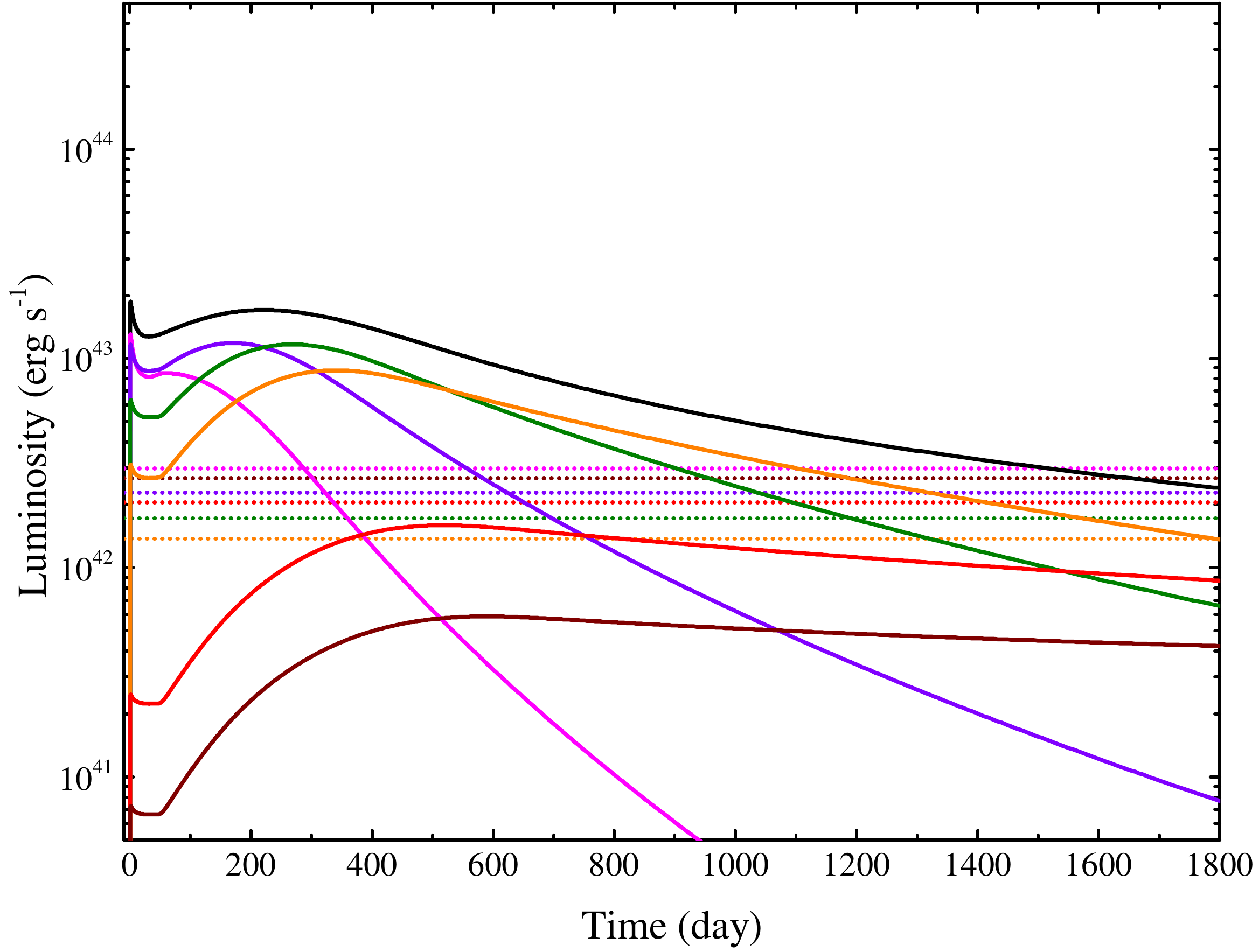} \\
\includegraphics[width=0.32\hsize]{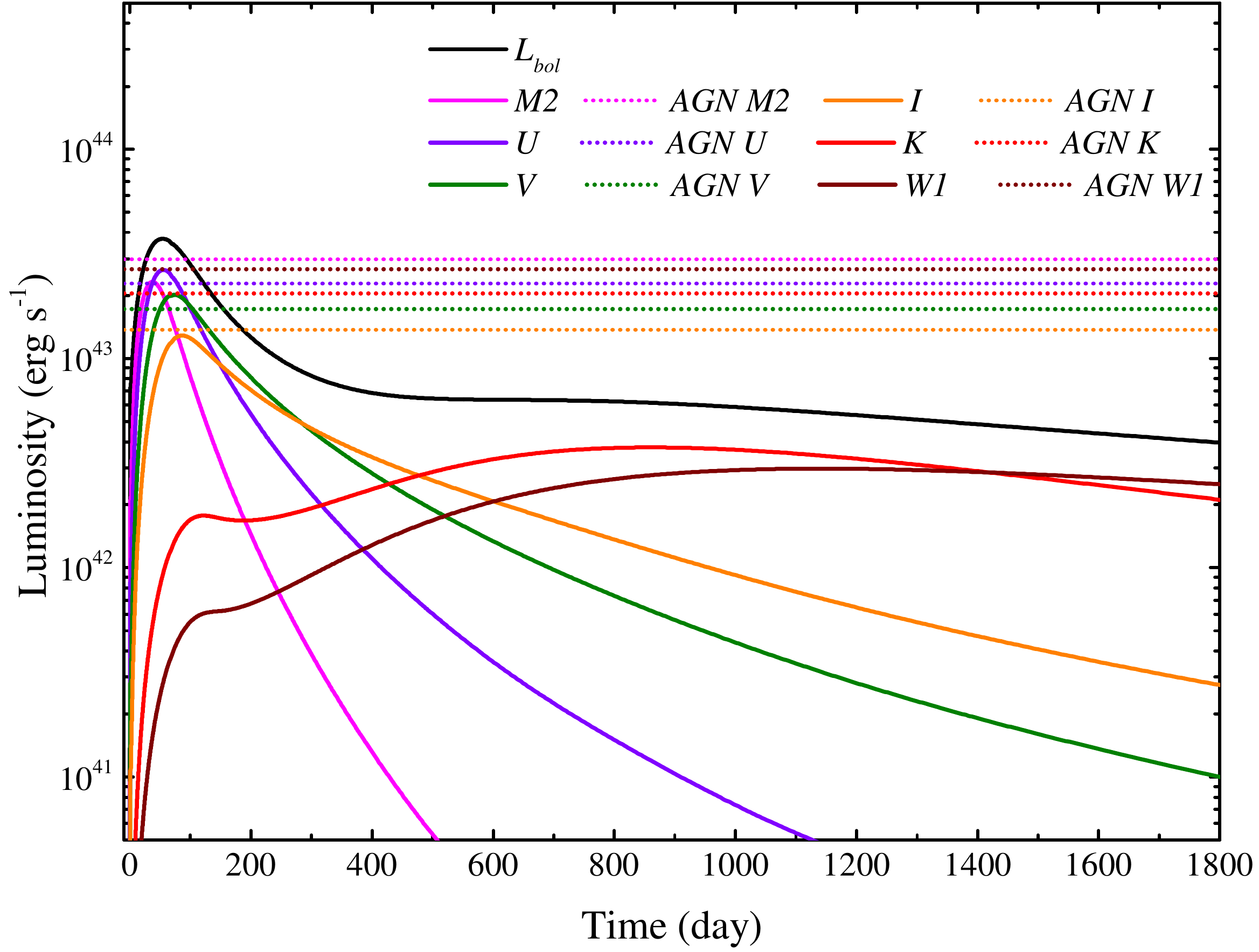}
\includegraphics[width=0.32\hsize]{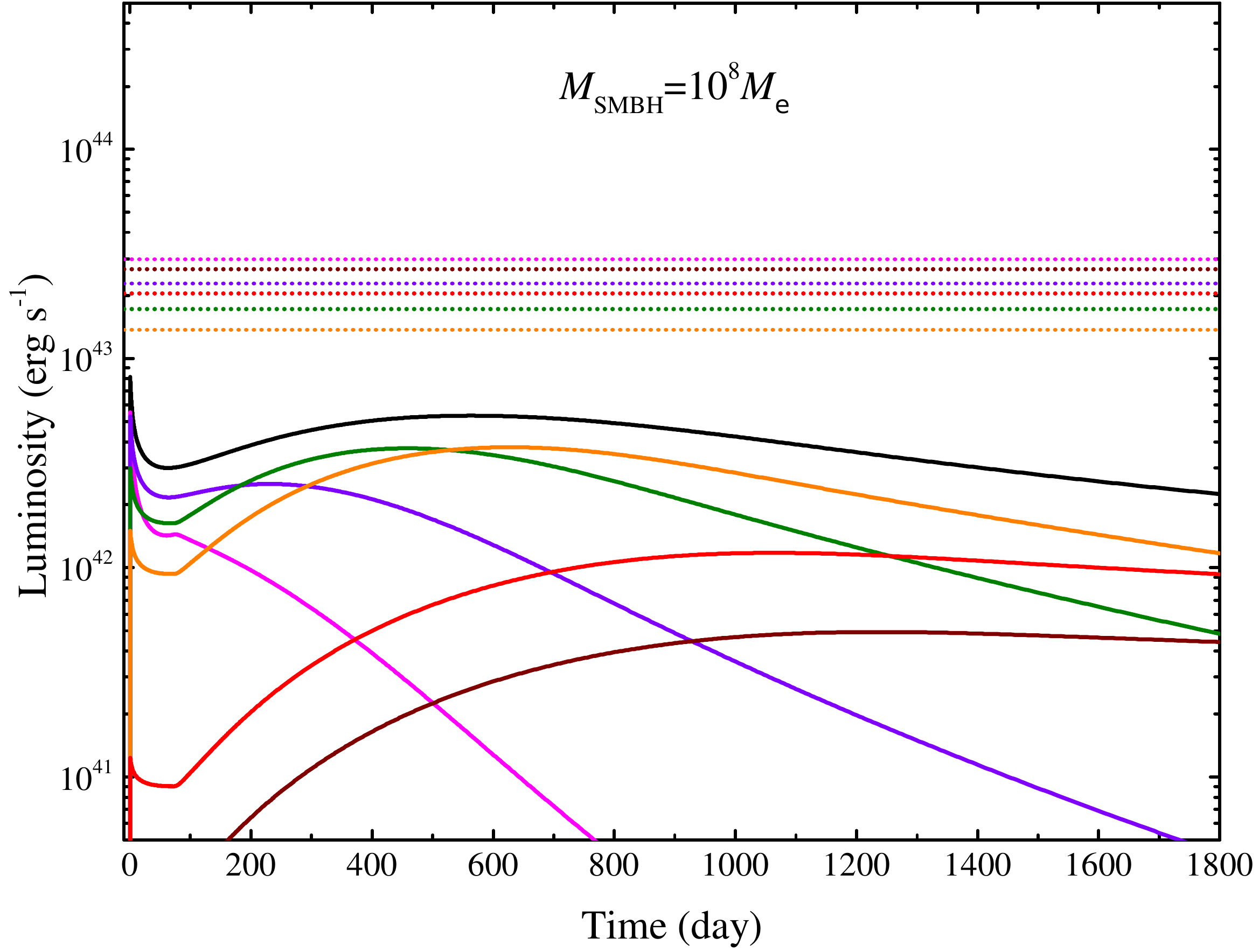}
\includegraphics[width=0.32\hsize]{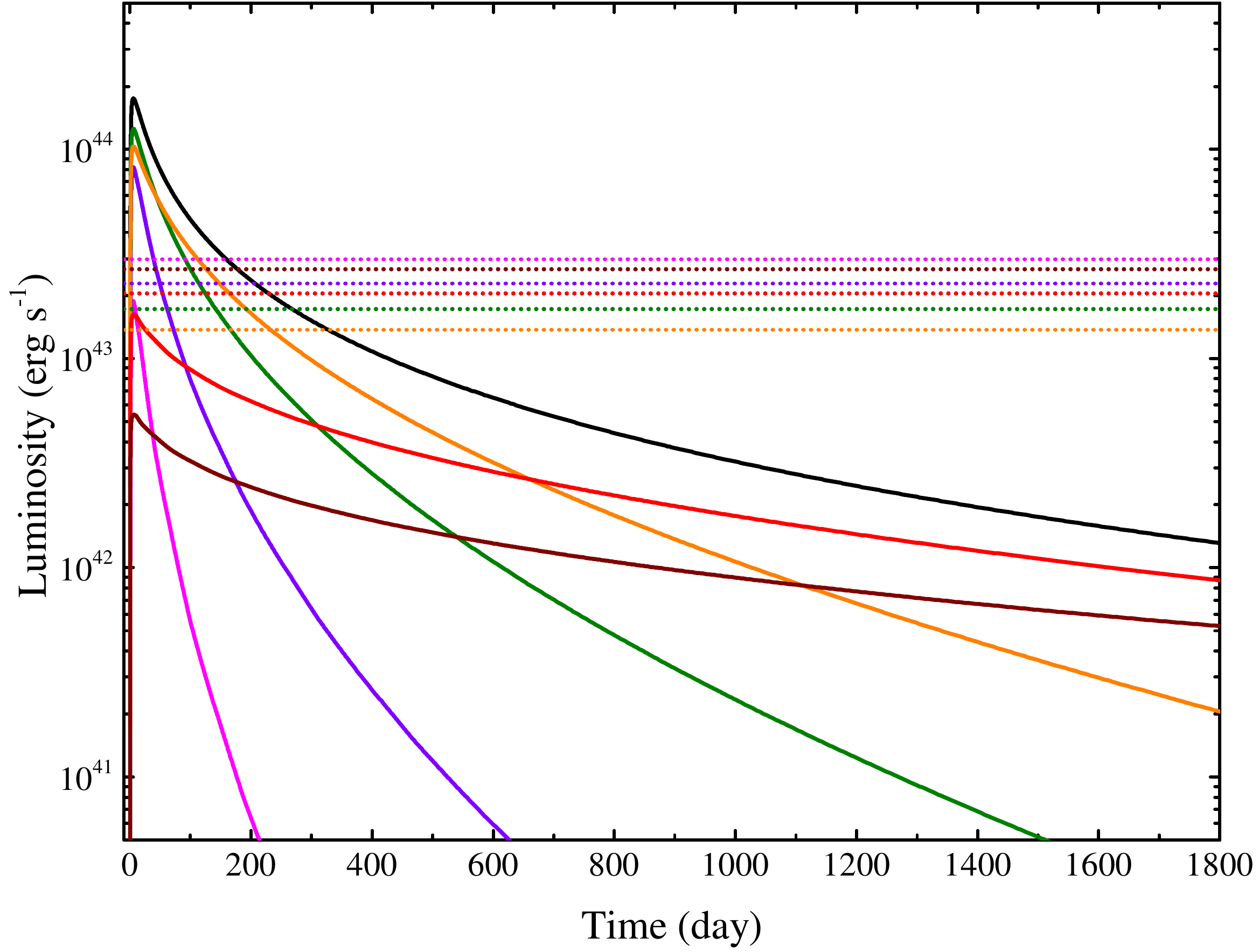}
\caption{The bolometric and multiband $\nu L_{\nu}$ light curves,
where \emph{M2} is \emph{UVOT/M2}, and \emph{W1} is \emph{WISE/W1}, respectively.
From top to bottom, the SMBH mass is $10^6$, $10^7$, and $10^8M_{\odot}$, respectively.
Panels from left to right hand sides show the results for mergers occurring at
$10^2$, $10^3$, and $10^4r_{\rm s}$ of the AGN disk, respectively.
the accretion rate of SMBHs is $\dot{M}_{\bullet}=0.1\dot{M}_{\bullet,\rm Edd}$,
other parameters can be found in Section~{\ref{results}}.
The horizontal lines show the background $\nu L_{\nu}$ luminosities of AGN at different bands.
We take the mean spectral energy distribution (SED) of Quasars from \cite{Shang_2011_Brotherton_apjs_v196.p2..2},
with the use of scaling law $\zeta$(3000\AA)=5.2 in the correlation of $L_{\rm bol,iso}=\zeta\lambda L_{\lambda}$
\citep{Runnoe_2012_Brotherton_mnras_v426.p2677..2688}.
}
\label{MyFig2}
\end{figure*}

The bolometric and multiband $\nu L_{\nu}$ luminosity curves
based on the numerical solutions are presented in Figure~{\ref{MyFig2}}.
It is no doubt that the bolometric light curves of our result
are brighter and longer lasting than the typical kilonova.
The radioactive decay energy powered by the $r$-process
elements is only a small fraction of the total energy that can be radiated.
Similar to the interacting supernova, most of the radiation energy is produced
by the interaction between ejecta and circumburst density gas.
Considering this fact, we call the transient an ``interacting kilonova" (IKN).

The peak luminosity of an IKN can be brighter than ${10^{44}~\rm erg~s^{-1}}$,
with similarity to superluminous supernovae (SLSNe) and tidal disruption events (TDEs).
So the IKN could be one of the most energetic stellar optical transients in the universe.
Note that IKN occurs in the central region of a galaxy,
and thus it most likely forms a part of the observed population of nuclear transients.
In Figure~\ref{MyFig2},
we also plot the background $\nu L_{\nu}$ luminosity of AGN at each band.
One can observe that the UV-optical-IR emission of an IKN can exceed the AGN background.
This makes an IKN become a very promising EM counterpart of GW events in the AGN disk for observation.

The multiband $\nu L_{\nu}$ luminosity curves
show the high-to-low evolution of peak radiation frequency over time.
This indicates the trend of evolution for the location and temperature of radiation regions.
With an increase in $\theta$ angle, radiation peak times of ejecta in different directions get delayed.
Ejecta with a small $\theta$ angle is likely to break out quickly
on account of their short propagation path to the disk surface.
This part of the ejecta becomes optically thin first,
releasing bright, and short-duration radiation that is biased towards the UV band
due to their relatively small photosphere radii.
Meanwhile, a larger $\theta$ angle means more disk gas that is swept,
which lowers the velocity of ejecta and makes it more difficult for photons to escape.
Both of these effects imply a lower peak radiation flux
\citep{Arnett_1982__apj_v253.p785..785}.
A later escape time of photons also implies that
the shell expands further before it is optically thin,
so a larger photosphere radius can be expected.
So, the effective temperature in the radiation area drops sufficiently,
and the peak radiation frequency shifts to the infrared band.

It takes approximately ten to twenty days for the UV band emission to reach its peak,
while the optical band peaks about thirty to fifty days later.
In most cases, the UV-optical emission could last tens to hundreds of days after the merger,
before they are dimmer than the background emission.
The rising time of IR emission is longer than that of UV-optical emission.
This time ranges from one hundred days to hundreds of days.
Generally, it follows the principle that the more massive the SMBH is,
the further the merger event takes place from the SMBH, and the longer the IR emission rises.
It means that the disk height at the merging position has a significant effect on IR emission.
This is in accordance with what we conclude from the dynamics,
which indicates that the radiation behavior of ejecta is determined by its travel time in the disk.
If the merger occurs in the outer disk, the dominant component of emission will be optical,
while if the merger occurs in the inner disk, the dominant component will be ultraviolet.
Based on these properties,
it is possible to estimate where the transient occurred in the disk of the AGN.

IKN bolometric luminosity decreases with increasing
binary merger radius for the same AGN disk parameters.
It is easy to imagine that this is due to the disk density decreasing with the radius increase.
One can also see this phenomenon between different masses of SMBHs.
Interestingly, an exception can be seen in the last panel of Figure~{\ref{MyFig2}},
with the SMBH mass being $10^8M_{\odot}$,
showing the effect of gas opacity on the disk.
While in the outermost disk parts, the temperature and density of disk drop,
the gas reaches the ``opacity gap'' at low temperature
($T\sim 10^3-10^4 ~\rm{K}$, see Figure 1 of \citealp{Thompson_2005_Quataert_apj_v630.p167..185}).
Due to the reduced optical depth, it is easier for photons to escape,
resulting in the radiation behavior being more like a typical interacting SLSN
(e.g., \citealp{Chatzopoulos_2012_Wheeler_apj_v746.p121..121}).

Especially, when the merger occurs at $10^2-10^3r_{\rm s}$
on the AGN disk with SMBH mass of $10^6 M_{\odot}$,
the UV-optical band appears to be re-brightened.
AGN disk properties as well as the ejecta geometry effect are responsible for this phenomenon.
The density of the accretion disk is highest in the middle zone,
and relatively low in the inner and outer zones.
Given the propagation of ejecta in different directions,
ejecta with larger $\varphi$ angles experiences
a decrease-increase-decrease disk environment of gas density,
which results in a re-brightening of the lightcurve at a later time than the peak.
Among more massive SMBHs,
this effect is masked by the bright radiation from ejecta in other $\varphi$ directions.
A large cavity will be formed in the AGN disk as a result of the expansion of the IKN-like stellar explosion.
The presence of such an event in the disk near the central SMBH
may cause the luminosity of the AGN itself to decline for some time following the event.

\begin{figure*}
\centering
\includegraphics[width=0.32\hsize]{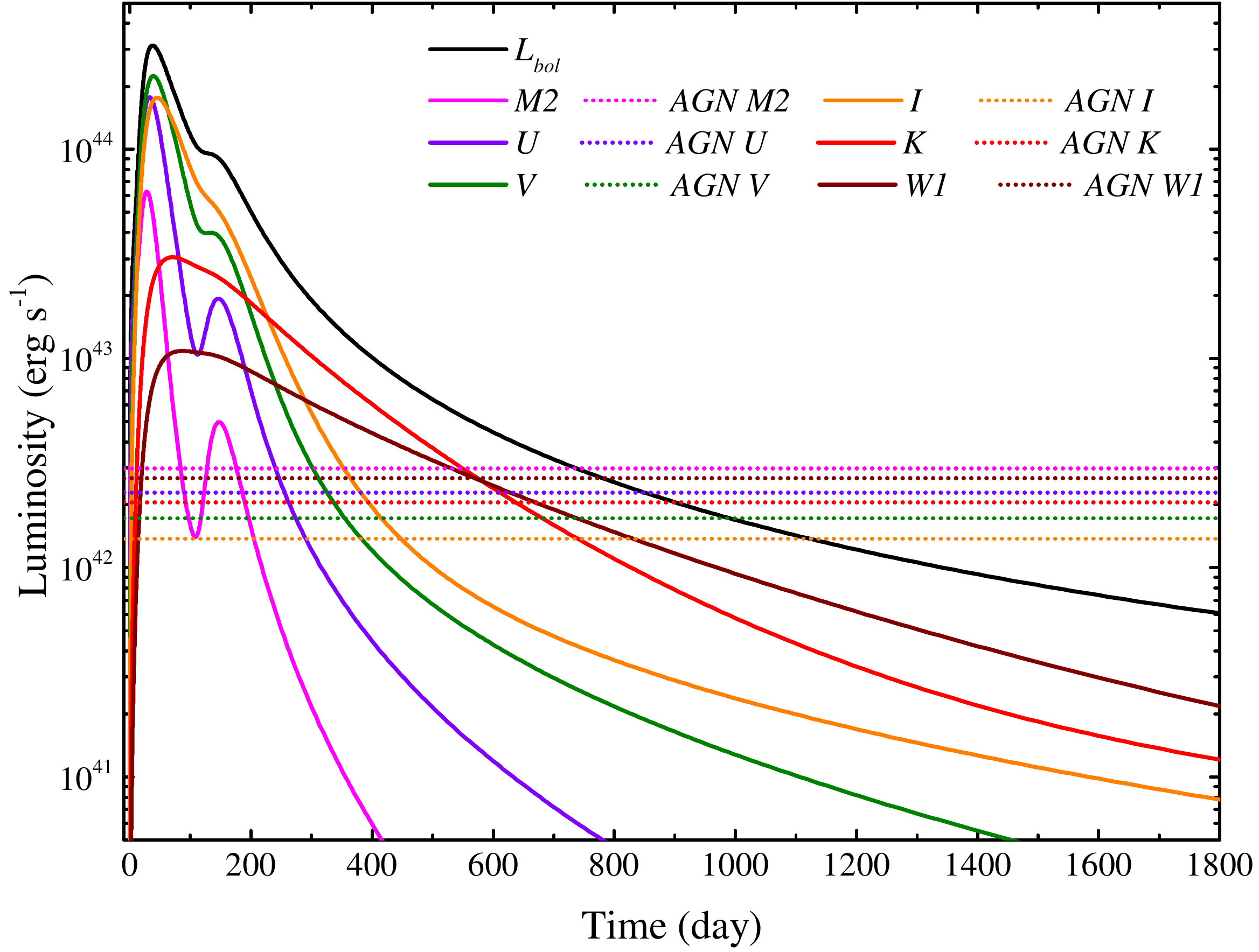}
\includegraphics[width=0.32\hsize]{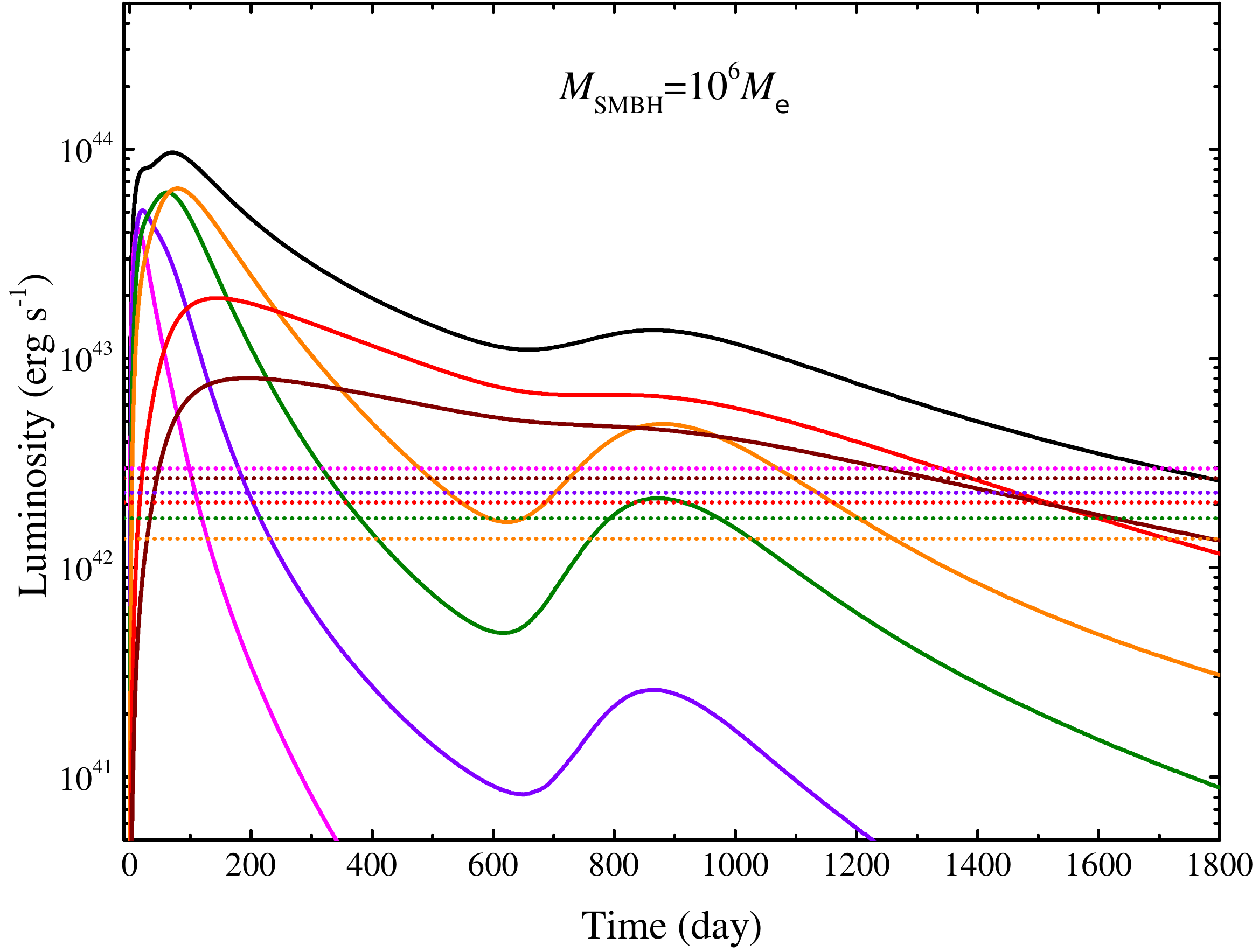}
\includegraphics[width=0.32\hsize]{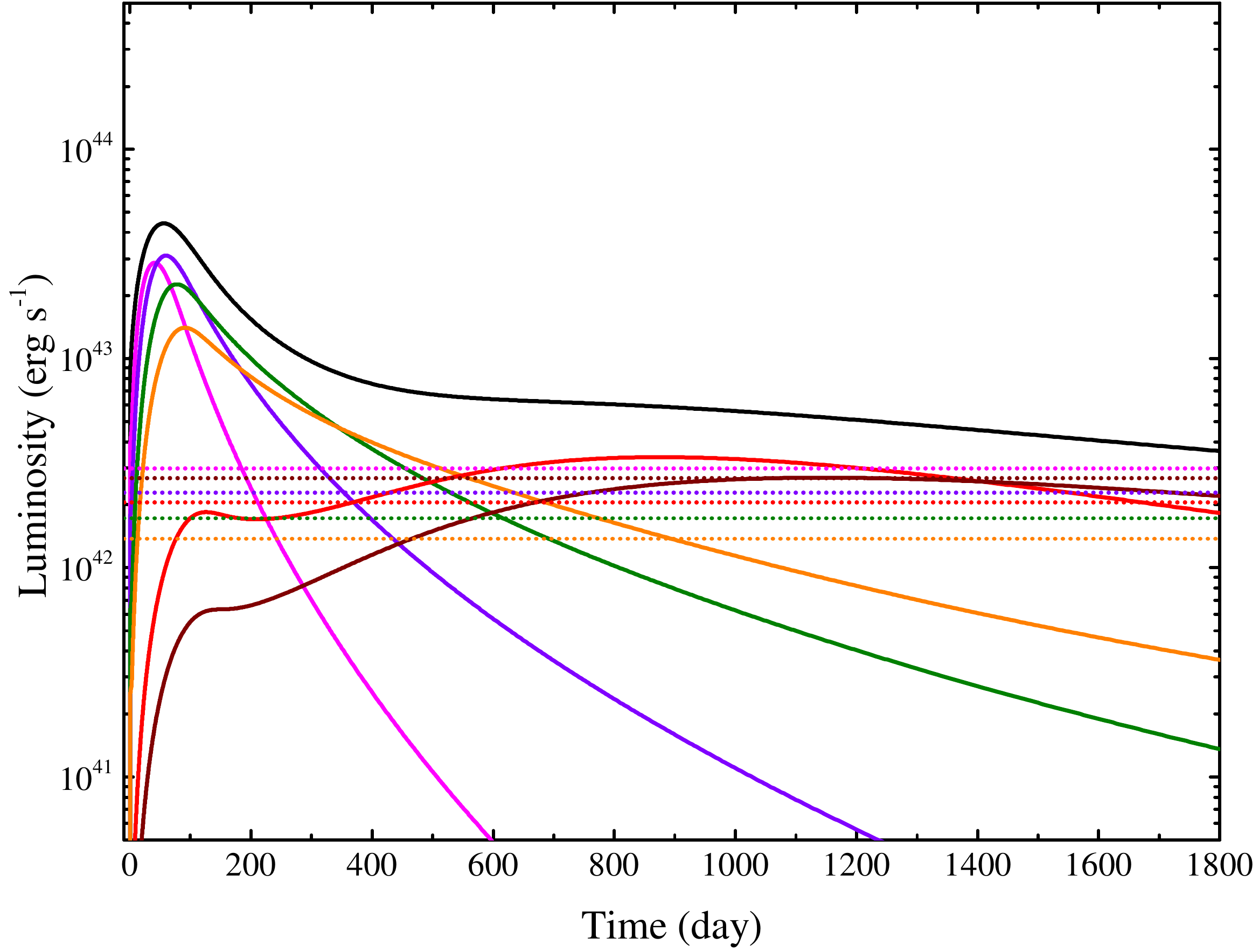} \\
\includegraphics[width=0.32\hsize]{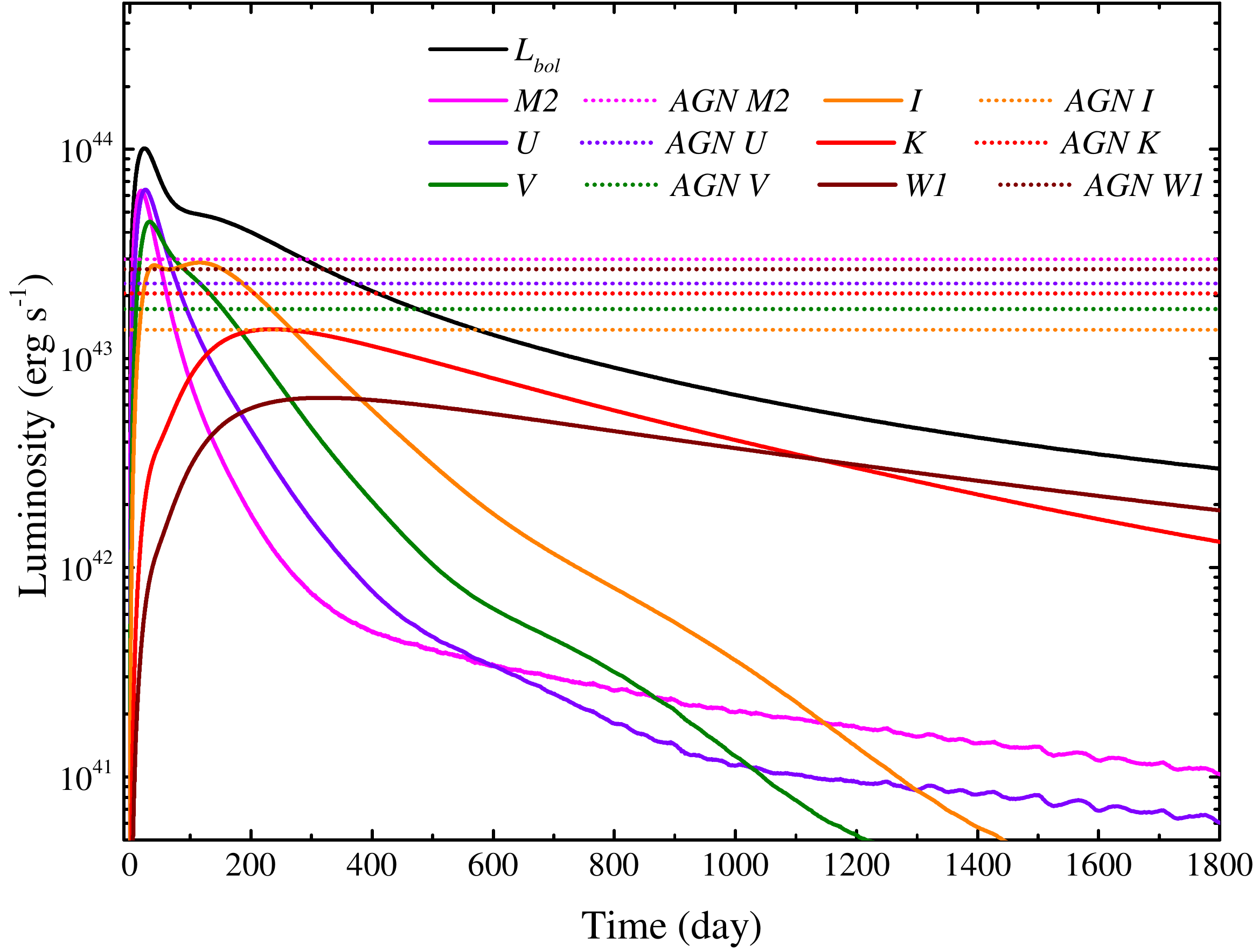}
\includegraphics[width=0.32\hsize]{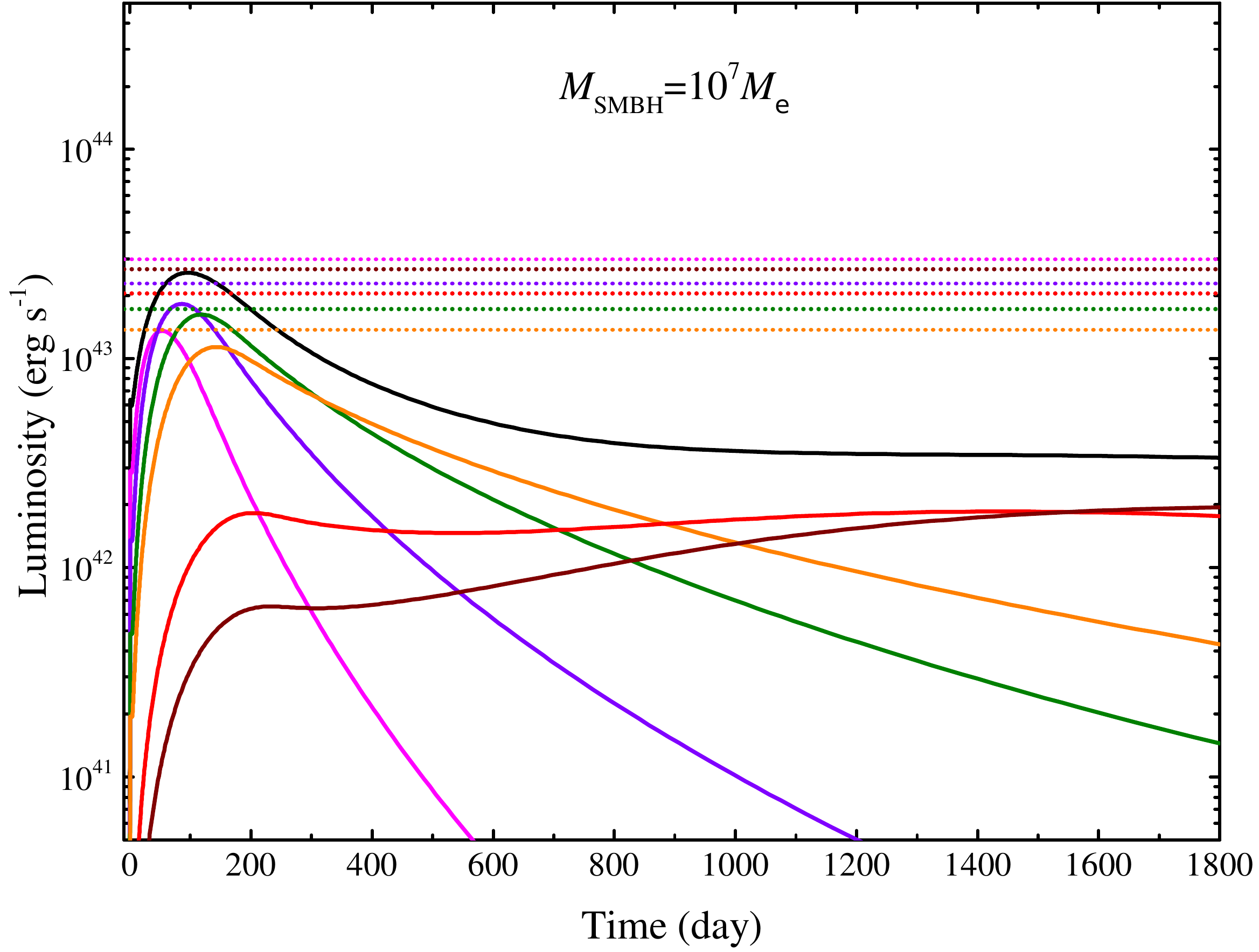}
\includegraphics[width=0.32\hsize]{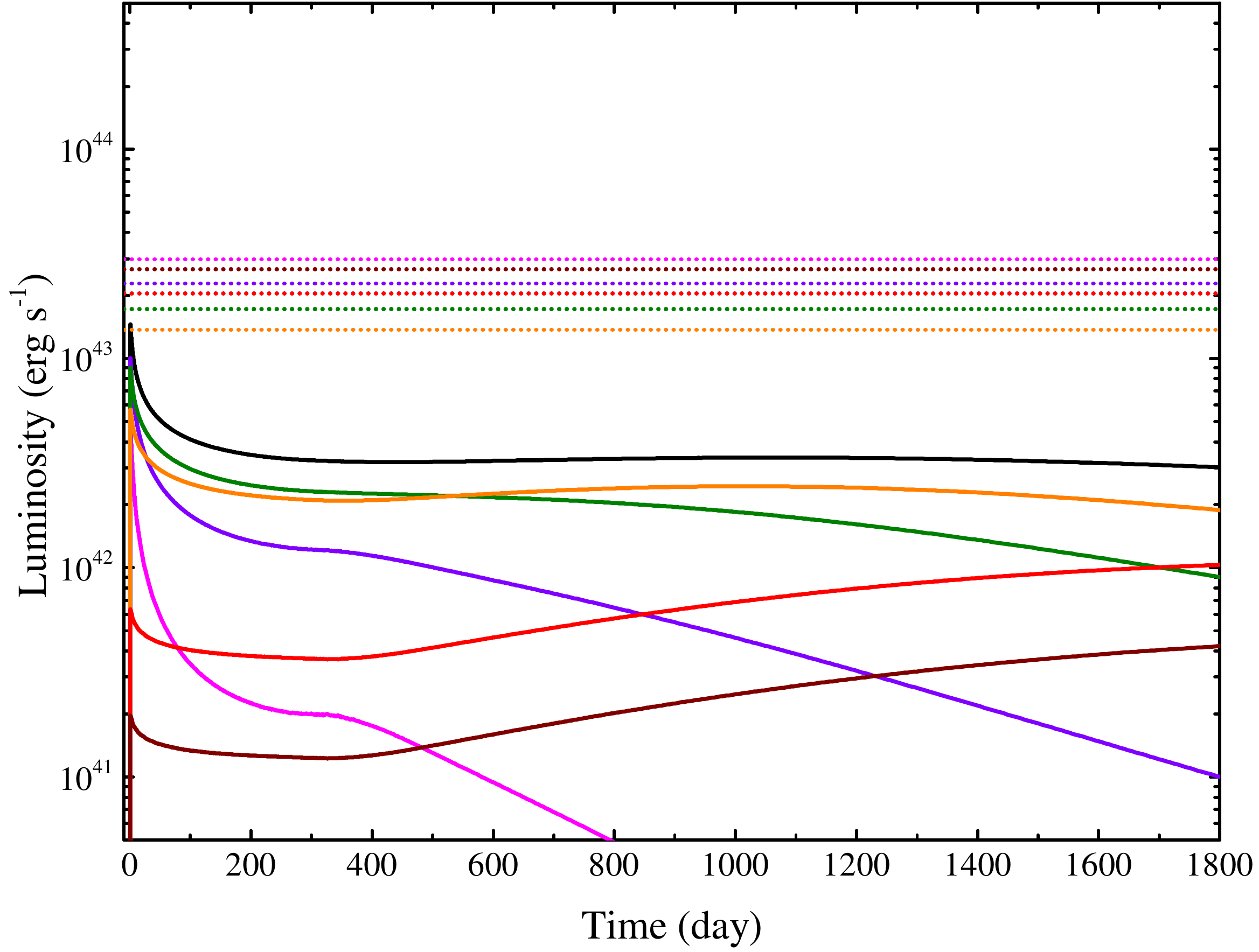} \\
\caption{Same as Figure~\ref{MyFig2}, but for SMBHs with accretion rate
$\dot{M}_{\bullet}=\dot{M}_{\bullet,\rm Edd}$.
}
\label{MyFig3}
\end{figure*}

\begin{figure}
\centering
\includegraphics[width=0.99\hsize]{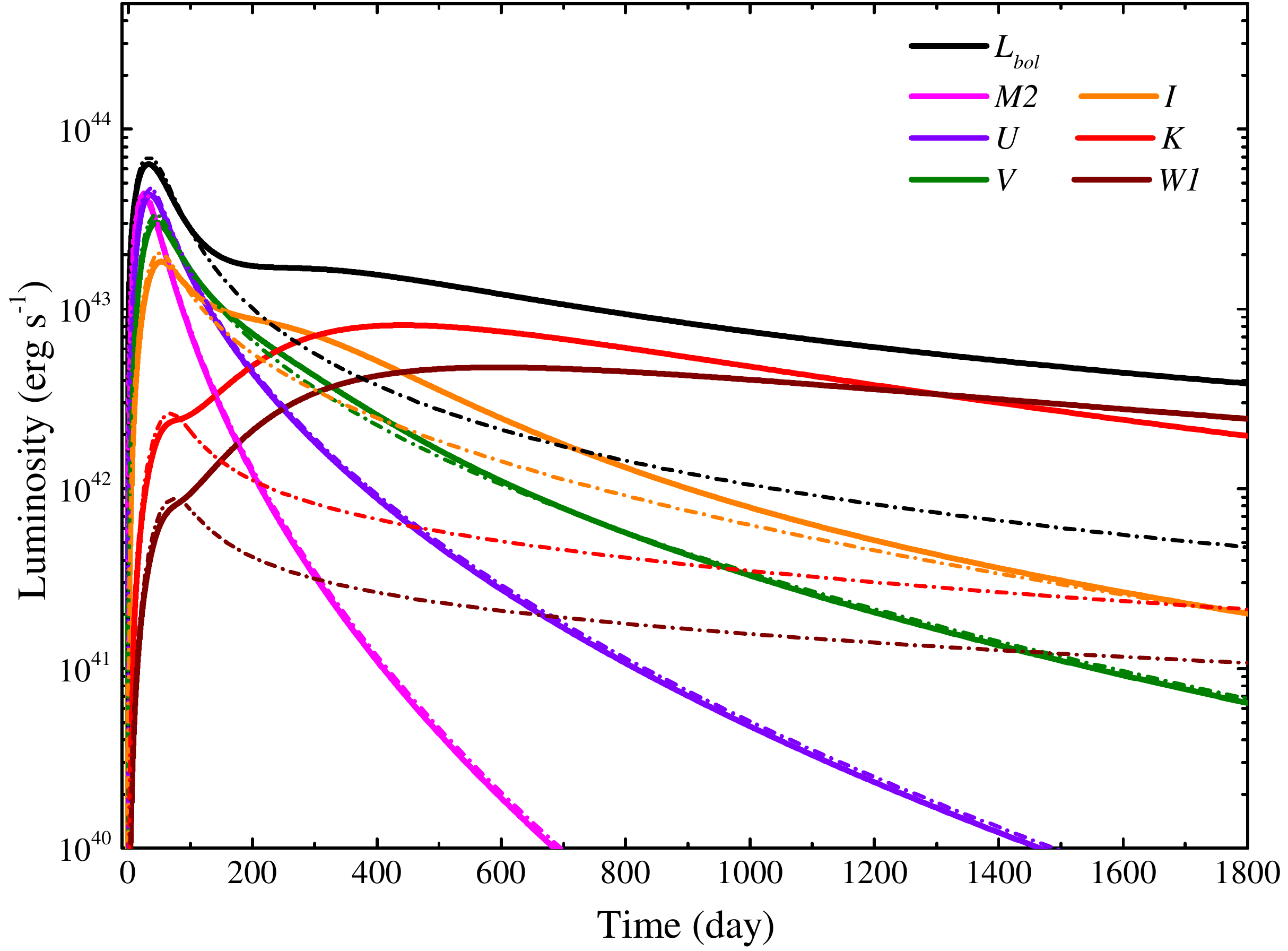}
\caption{The lightcurves under different AGN nucleus region environments,
with gas densities of $10^{-24}~\rm g~cm^{-3}$ (dash-dotted lines)
and $10^{-17}~\rm g~cm^{-3}$ (solid lines), respectively.
The mass of SMBH is $10^7~M_{\odot}$ with accretion rate
$\dot{M}_{\bullet}=0.1\dot{M}_{\bullet,\rm Edd}$,
merger has occurring at $10^3r_{\rm s}$ of the AGN disk,
other parameters are the same as Figure~\ref{MyFig2}
and fixed.
}
\label{MyFig4}
\end{figure}

\section{discussion}\label{discuss}
\subsection{The effect of GRB jet}
After the binary merger, a pair of jets may be produced.
Radiation produced by the jet in the AGN disk environment has been discussed.
We note that the possible shock breakout signal \citep{Zhu_2021_Zhang_apj_v906.p11..11L},
the GeV to sub-TeV radiation \citep{Yuan_2022_Murase_apj_v932.p80..80},
and the long-lasting modified afterglow \citep{Wang_2022_Lazzati_mnras_v.p..}
for a jet may be the precursors to or analogs of an IKN.
Although these processes require different environments
(a cavity formed or not) in the progenitor system.

The jet will likely be choked if it propagates in a dense environment
\citep{Perna_2021_Lazzati_apjl_v906.p7..7L}.
A jet that maintains a small half-opening angle ($\sim 10^\circ$)
and does not expand laterally will not significantly affect the structure of the disk,
so there will be no significant change about the subsequent IKN light curve.
In contrast, if the jet is substantially obstructed, there is an obvious lateral expansion.
Therefore, the jet could undergo quasi-spherically symmetric expansion in the disk,
sweep up more gas within the larger half-opening angle,
and break out from the disk at a lower velocity.
The shock breakout emission is reduced
in both luminosity and peak radiation frequency in this case.
Therefore, the situation would differ from that described by
\cite{Zhu_2021_Zhang_apj_v906.p11..11L},
who only discussed the scenario of a non-lateral expanding jet.
We believe that this process can also affect the early radiation of an IKN
because the gas swept up in advance will result in a weakening of
the interaction of the gas and ejecta.

\subsection{The influence of different accretion rates of an AGN disk}
AGNs have a broad range of luminosity reaching up to the high-Eddington luminosity
\citep[e.g.,][]{Simmons_2012_Urry_apj_v761.p75..75}.
Figure~\ref{MyFig3} illustrates the IKN lightcurves when the accretion rate of SMBH increases to
$\dot{M}_{\bullet}=\dot{M}_{\bullet,\rm Edd}$.
We have found that an IKN is difficult to detect from
the background emission of SMBH with a mass of $10^8 M_{\odot}$
and high-Eddington luminosity, so we do not depict this situation in the figure.
Our results indicate that the increased accretion rate of SMBH
does not have a significant effect on IKN lightcurves.
Observations are adversely affected by the brightening background radiation
that renders it more difficult to detect an IKN.
In summary, the results of Figures~\ref{MyFig2} and~\ref{MyFig3}
suggest that an IKN occurring when an AGN has a lower-mass SMBH
is more likely to be observed because a larger parameter space is allowed.

\subsection{The influence of gas density in a nuclear sphere}\label{density}
The nuclear sphere of AGN has a complex gas environment
(e.g., \citealp{Revalski_2022_Crenshaw_apj_v930.p14..14}).
We examine here how the gas environment affects the radiation behavior of an IKN.
For simplicity, we have not considered the radius and structure of gas in the nuclear sphere.
We have compared the results for gas densities of
$10^{-24}~\rm g~cm^{-3}$ and $10^{-17}~\rm g~cm^{-3}$ with all the other parameters fixed.
The result is shown in Figure~{\ref{MyFig4}}.
One can observe that the gas density of the nuclear sphere primarily affects
the late IR band light curves, but has little effect on the early UV-optical radiation.
An IKN will no longer be able to emit bright, long-lasting IR emission
in a very empty nuclear sphere environment.
This feature will help to diagnose the environment in which IKN occurs.
Moreover, this result supports the view that the early UV-optical radiation
is mainly due to the cooling process which occurs after the ejecta breaks out of the disk.

\subsection{IKN from the merger of a misaligned binary}
For compact binary populations within the AGN nuclear region,
\cite{Tagawa_2020_Haiman_apj_v898.p25..25}
found that approximately twenty percent of mergers occur within the disk of the AGN.
There are some embedded binaries with orbital angular momentums
that are not aligned with the AGN disk \citep{Fabj_2020_Nasim_mnras_v499.p2608..2616};
however, some other works have shown that the orbital angular momentum
is expected to be aligned or antialigned with the AGN disk
\citep{McKernan_2018_Ford_apj_v866.p66..66,McKernan_2020_Ford_mnras_v494.p1203..1216,
Secunda_2019_Bellovary_apj_v878.p85..85,Yang_2019_Bartos_prl_v123.p181101..181101}.
Note that the in-situ compact stars on the AGN disk are naturally aligned with the disk.
Whatever, as shown in Section~{\ref{density}},
the early UV-optical radiation mainly comes from the cooling process after the ejecta breaks out of the disk.
Consequently, faster and more energetic ejecta are driving brighter UV-optical peak radiation.
Therefore, the misaligned merger in the disk may result in brighter and shorter-duration radiation.

\subsection{The event rate of IKNe}
Ejecta that expands in all directions enhances the probability of its being detected.
Considering that the observed merger rate of BH-NS/NS-NS binaries with remnant ejecta is
${0.1R_{\rm BH-NS}+R_{\rm NS-NS} \sim (f_{\rm AGN}/0.1)\left[0.12,43\right] ~\rm{Gpc}^{-3}~\rm{yr}^{-1}}$
\citep{McKernan_2020_Ford_mnras_v498.p4088..4094},
IKNe can be detected with a similar rate in the best-case scenario.
However, it is expected that some IKNe cannot exceed the background of AGN radiation
under the low kinetic energy of ejecta and/or high luminosity of AGNs.
Additionally, the merger rates of the BH-NS and NS-NS binaries
depend on the specific properties of the SMBH and the AGN disk.
Confirmation of this ratio requires further research, which is beyond the scope of this paper.

The origin of AGN transient sources is still unclear.
Considering that the kinetic energy of IKN is $\sim 10^{51}$~erg,
some transients with total radiation energy less than
this value are more likely to be candidates for IKNe.
In terms of event rates,
\cite{Sun_2015_Zhang_apj_v812.p33..33}
found that the TDE event rate densities at different peak luminosity bins
have a wide distribution ranging from $\sim 10^4~\rm Gpc^{-3}~yr^{-1}$ at
$10^{43}~\rm erg~s^{-1}$
to $\sim 10^2~\rm Gpc^{-3}~yr^{-1}$ at $10^{45}~\rm erg~s^{-1}$.
In contrast, the event rate of IKNe is at least two orders of magnitude lower.
\cite{Assef_2018_Prieto_apj_v866.p26..26}
indicated that IR-bright nuclear transients in the AGN
detected by \emph{WISE} (Wide-field Infrared Survey Explorer)
have a lower limit on the event rate as $>1.2~\rm Gpc^{-3}~yr^{-1}$.
This event rate can be satisfactory for IKNe
even if we assume that the IKNe with bright IR radiation
is only one percent of the population.
Besides, the event rate of type~I hydrogen-poor SLSNe is
$68^{+94}_{-44}~\rm Gpc^{-3}~yr^{-1}$ in our local universe
and is $151^{+151}_{-82}~\rm Gpc^{-3}~yr^{-1}$ for type~II hydrogen-rich SLSNe
\citep{Quimby_2013_Yuan_mnras_v431.p912..922}.
Since a portion of the AGN transients is interpreted as SLSNe,
so IKNe may also be included as a component of the population.
Certainly, the most direct evidence is the association of an IKN with a GW event,
which provides an explicit estimate of the event rate of IKNe.

\section{Summary}\label{Summary}
In this work, we have investigated the dynamics and EM signatures of
the NS-NS/NS-BH merger ejecta that occur in an AGN disk.
We proposed a three-dimensional, self-consistent thin shell model,
but ignored the interaction of ejecta between different angles,
to describe the dynamics and radiation properties of ejecta.
We found that the interaction between ejecta and disk gas
produces a profound effect on the behavior of dynamics and radiation.
For a typical dynamic process of ejecta, five stages need to be accomplished:
(1) gravity-induced slowing down, (2) coasting, (3) Sedov-Taylor deceleration in the disk,
(4) re-acceleration after the break out from the disk surface,
and (5) momentum conserved snowplow phase in the galactic nucleus region.
The specific dynamics depend on the angle of the ejecta
and the properties of the disk and ejecta themselves.
Besides, the radiation from the ejecta is so bright
that its peak luminosity reaches a few $10^{43}-10^{44}~\rm erg~s^{-1}$.
In extreme cases, a more dense disk environment and energetic merger ejecta
could produce a transient with peak luminosity of $\sim 10^{45}~\rm erg~s^{-1}$.
As a result, it is one of the brightest stellar optical transients in the universe,
comparable to the SLSNe and TDEs
that probably occur also in the nuclei of galaxies.
Because most of the radiated energy has transformed from the kinetic energy of merger ejecta,
thus we call the transient interacting kilonova (IKN).

IKN is a promising bright EM counterpart to NS-NS/BH-NS merger event in the AGN disk.
We found that for the sub-Eddington accretion AGN disk ($0.1L_{\rm Edd}$)
with relatively less massive SMBHs
($<10^8M_{\odot}$, most of SMBHs in the universe\footnote{
\url{http://www.astro.gsu.edu/AGNmass/}},
\citealp{Bentz_2015_Katz_PASP_v127.p67..73}),
IR-optical-UV radiation from IKN exceeds that from AGN themselves.
In addition, we plotted the results for a high-Eddington accretion AGN disk ($1L_{\rm Edd}$)
and found that optical-UV emission from IKN
could exceed the background emission from the AGN disk
when IKN occurred in the disk of $M_{\rm SMBH}<10^7M_{\odot}$.
Based on this result, IKNe occurring in low-mass SMBH AGNs
are more straightforward to observe due to the larger parameter space allowed.

The peak radiation of an IKN occurs in the UV band first
and gradually moves to the lower energy band due to the evolution.
It takes approximately ten to twenty days for the UV band emission to reach its peak,
while the optical band peaks about thirty to fifty days later.
The long rising time and bright peak luminosity give most survey telescopes ample time to search.
In most cases, the UV-optical emission could last tens to hundreds of days after the merger,
before they are dimmer than the background emission.
The rising time of IR emission is longer than that of UV-optical emission,
ranging from one hundred days to hundreds of days.
Overall, the IR emission can keep bright for thousands of days.
This means that it is easy to search for the EM counterpart of the GW event
that occurs in the AGN disk.
However, the similar peak brightness, peak time,
and evolution pattern of the light curve of IKN and SLSNe/TDEs
make it difficult to tell the difference between them.
But on the other hand,
it also means that IKN is likely to have been present in recorded AGN nuclei transients.
For example, it may be related to some of the AGN transients
that are IR-bright with ultra-long durations
(e.g., \citealp{Assef_2018_Prieto_apj_v866.p26..26,Reynolds_2022_Mattila_aap_v664.p158..158A}).

It should be noted that we made a series of simplifications to the model in this work.
For instance, the radial structure of the ejecta, kick velocity of the compact remnant,
fallback accretion, and the effect of the AGN disk environment are not considered.
We also simplified the radiative transfer process and ignored the influence of dust,
which may lead to a too-idealistic radiation behavior.
Further numerical simulations will provide more detailed information
about the dynamics, nucleosynthesis, and radiation of IKNe.

\acknowledgments
We thank the anonymous referee for helpful comments and suggestions.
RJ thanks Da-Bin Lin and Lu-Yao Jiang for helpful discussions.
RJ also thanks Guo-Peng Li for sharing SG disk codes.
This work was supported by
the National Key Research and Development Program of China (grant No. 2017YFA0402600),
the National SKA Program of China (grant No. 2020SKA0120300),
the National Natural Science Foundation of China (grant No. 11833003),
RJ acknowledges support by
the Postgraduate Research \& Practice Innovation Program of Jiangsu Province  (grant No. KYCX22\_0105).


\software{\texttt{Matplotlib}
 \citep{Hunter_2007__ComputinginScienceandEngineering_v9.p90..95},
 \texttt{Numpy}
 \citep{Harris_2020_Millman_Nature_v585.p357..362}}



\end{document}